\documentclass[lettersize,journal]{IEEEtran}
\usepackage{amsmath,amsfonts}
\usepackage{algorithmic}
\usepackage{algorithm}
\usepackage{array}
\usepackage{textcomp}
\usepackage{stfloats}
\usepackage{url}
\usepackage{verbatim}
\usepackage{graphicx}
\usepackage{cite}
\usepackage{lipsum}
\usepackage{siunitx}
\usepackage{subcaption}
\usepackage{multirow}
\usepackage[table,xcdraw]{xcolor}
\usepackage[normalem]{ulem}
\usepackage{amssymb}
\usepackage{pifont}

\newcommand{\xmark}{\ding{55}}%

\begin{document}

\title{A Heterogeneous RISC-V based SoC for Secure Nano-UAV Navigation}

\author{
\IEEEauthorblockN{
Luca Valente, 
Alessandro Nadalini, 
Asif Veeran, 
Mattia Sinigaglia, 
Bruno Sá, 
Nils Wistoff,~\IEEEmembership{Student Member,~IEEE,} 
Yvan Tortorella, 
Simone Benatti, 
Rafail Psiakis,
Ari Kulmala,
Baker Mohammad,~\IEEEmembership{Sr. Member,~IEEE,} 
Sandro Pinto,
Daniele Palossi,
Luca Benini,~\IEEEmembership{Fellow,~IEEE,}
Davide Rossi,~\IEEEmembership{Member,~IEEE} 
} 
\thanks{Manuscript submitted July 30, 2023.}
\thanks{Luca Valente, Mattia Sinigaglia, Alessandro Nadalini, Yvan Tortorella, Simone Benatti, Luca Benini, and Davide Rossi are with the Department of Electrical, Electronic and Information Engineering, University of Bologna, 40136 Bologna, Italy. 
Asif Veeran and Baker Mohammad are with the Department of Electrical Engineering and Computer Science, Khalifa University, Abu Dhabi, UAE.
Bruno Sá and Sandro Pinto are with Centro ALGORITMI, University of Minho, 4800-058 Guimarães, Portugal. 
Nils Wistoff, Daniele Palossi and Luca Benini are with the Integrated Systems Laboratory (IIS), ETH Zürich, 8092 Zürich, Switzerland.
Ari Kulmala and Rafail Psiakis are with Technology Innovation Institute, Secure Systems Research Center, Abu Dhabi, UAE.
Daniele Palossi is also with the Dalle Molle Institute for Artificial Intelligence (IDSIA), USI-SUPSI, 6900 Lugano, Switzerland.}
\thanks{This work was supported by Technology Innovation Institute, Secure Systems Research Center, Abu Dhabi, UAE, PO Box: 9639, and by the Spoke 1 on Future HPC of the Italian Research Center on High-Performance Computing, Big Data and Quantum Computing (ICSC) funded by MUR Mission 4 - Next Generation EU, and by KDT TRISTAN project (g.a. 101095947)}
}

\markboth{}%
{Shell \MakeLowercase{\textit{et al.}}: A Sample Article Using IEEEtran.cls for IEEE Journals}


\maketitle
\bstctlcite{IEEEexample:BSTcontrol}
\begin{abstract}
The rapid advancement of energy-efficient parallel ultra-low-power (ULP) $\mu$controllers units (MCUs) is enabling the development of autonomous nano-sized unmanned aerial vehicles (nano-UAVs). 
These sub-10cm drones represent the next generation of unobtrusive robotic helpers and ubiquitous smart sensors.
However, nano-UAVs face significant power and payload constraints while requiring advanced computing capabilities akin to standard drones, including real-time Machine Learning (ML) performance and the safe co-existence of general-purpose and real-time OSs.
Although some advanced parallel ULP MCUs offer the necessary ML computing capabilities within the prescribed power limits, they rely on small main memories ($<$1MB) and $\mu$controller-class CPUs with no virtualization or security features, and hence only support simple bare-metal runtimes.
In this work, we present Shaheen, a 9mm$^{\textbf{2}}$ 200mW SoC implemented in 22nm FDX technology.
Differently from state-of-the-art MCUs, Shaheen integrates a Linux-capable RV64 core, compliant with the v1.0 ratified Hypervisor extension and equipped with timing channel protection, along with a low-cost and low-power memory controller exposing up to 512MB of off-chip low-cost low-power HyperRAM directly to the CPU.
At the same time, it integrates a fully programmable energy- and area-efficient multi-core cluster of RV32 cores optimized for general-purpose DSP as well as reduced- and mixed-precision ML. 
To the best of the authors' knowledge, it is the first silicon prototype of a ULP SoC coupling the RV64 and RV32 cores in a heterogeneous host+accelerator architecture fully based on the RISC-V ISA.
We demonstrate the capabilities of the proposed SoC on a wide range of benchmarks relevant to nano-UAV applications including general-purpose DSP as well as inference and online learning of quantized DNNs.
The cluster can deliver up to 90GOp/s and up to 1.8TOp/s/W on 2-bit integer kernels and up to 7.9GFLOp/s and up to 150GFLOp/s/W on 16-bit FP kernels.
\end{abstract}


\newpage
\section{Introduction}\label{sec:intro}

\IEEEPARstart{T}{he} number of Internet-of-Things (IoT) devices and the spectrum of IoT applications are rapidly growing: from home automation, robotics, industrial gateways, and building automation to smart cities, digital signage, medical equipment, and more~\cite{iot_review}. 
In this context, nano-sized unmanned aerial vehicles (nano-UAVs) can be considered the ``ultimate'' IoT node, thanks to their ability to navigate, sense, analyse, and understand the surrounding environment.
Nano-UAVs have a form factor of a few centimeters in diameter, and a weight of only tens of grams, which allows them to safely operate near humans and in narrow, cramped spaces~\cite{dronet,cereda2023deep}.
They have a total power envelope of a few Watts, of which only 5-15\% for computation~\cite{pico}, and their small physical footprint and limited payload restrain the maximum battery, the printed circuit board size and exclude any form of active cooling.
Nowadays, $\mu$controller units (MCUs) are the only computing platforms that meet the nano-UAV's power and form-factor constraints.
%

MCUs feature simple RISC host processors (e.g., ARM Cortex-M) with low computational capabilities and no virtualization support, to which they expose just a few hundred kBytes of on-chip SRAM scratchpad memory (SPM)~\cite{stm32h7, stm32f4, gap8, kraken, motion-control-isscc,navion,Ju}. 
To deliver more advanced computational capabilities, state-of-the-art (SoA) MCUs integrate accelerators with high data processing capabilities~\cite{stm32h7, stm32f4, gap8, kraken, motion-control-isscc,navion,Ju}. 
Usually, ultra-low-power (ULP) devices' accelerators are hardwired application-specific data-paths\cite{motion-control-isscc, navion} which achieve the best energy efficiency but are tailored to a single application domain, leading to poor programmability and a high nonrecurring engineering cost~\cite{dally_dsa} while occupying a considerable part of the scarce area resources.
To improve the overall versatility of the SoC, recent works replace ASIC accelerators with fully programmable parallel accelerators~\cite{gap8,kraken} that achieve competitive energy efficiency while maintaining significant flexibility, and hence make the most out of the available power and area.

The increase in computing capabilities of SoA MCUs has enabled nano-UAVs to achieve autonomous flight while executing intelligent auxiliary tasks.
For example, Quantized Neural Networks (QNNs) have been proposed to carry out obstacle avoidance \cite{dronet} or human pose estimation and object detection~\cite{cereda2023deep}.
At the same time, floating point (FP) digital-signal-processing (DSP) computation has been proposed for path planning or structural build monitoring~\cite{wavelet_drones,fft_uav}.
Also, a recent trend for edge devices is online learning, which enables a small portion of the Neural Networks (NN) training to happen on the edge, increasing the accuracy and reliability directly on the field~\cite{amodei2016concrete,trainlib_git}.
Nevertheless, even the most advanced MCUs, supporting this new class of applications, lag behind in terms of software support.
Due to the small amount of available memory and the simplicity of their host CPU, SoA MCUs only provide close-to-metal software environments, based on minimal real-time operating systems (RTOS) or simple bare-metal runtimes.
However, enabling the execution of full-fledged OSs (like Linux), securely along with the real-time control applications, would allow nano-UAVs to leverage an existing, mature, and solid software stack and hence ease the software development~\cite{agilicious}.
%
In this context, this work presents a step forward in the current and future generation of autonomous nano-UAVs.
We present Shaheen, a 9$mm^2$ 200mW heterogeneous System-on-Chip (SoC) implemented in \SI{22}{\nano\meter} FDX technology that couples an application-class RV64 host processor with a low-power HyperRAM memory controller and with a flexible cluster of eight RV32 cores, providing best-in-class energy efficiency and performance for IoT applications.
The design is fine-tuned to accommodate the requirements of emerging nano-UAV applications.

The host includes hardware virtualization support~\cite{bruno_h}: to the best of our knowledge, it is the first silicon implementation fully compliant with the ratified RISC-V Instruction Set Architecture (ISA) Hypervisor extension\footnote{SiFive, Ventana, and StarFive have announced RISC-V CPU designs with Hypervisor extension support, but we are not aware of any silicon available on the market yet.}, enabling the secure coexistence of an RTOS and a full-blown OS onto the same host core.
In particular, the Hypervisor extension aims to provide confidentiality and integrity of virtual machines (VM) by enforcing isolation (via two-stage virtual memory) between multiple consolidated guest OSes, i.e., General Purpose OS (GPOS) and RTOS.
To further isolate the execution of these coexisting software stacks (trusted and untrusted), prevent security threats, and ensure multi-domain operations, the host core features Physical Memory Protection (PMP)~\cite{pie} and ISA and micro-architecture extensions for timing channel mitigation~\cite{nils_fence}.
Namely, the PMP aims to provide confidentiality and integrity by limiting the physical addresses accessible by software running on CVA6. PMP enforces the separation between the bare-metal firmware (running in machine mode) and everything else through a set of additional registers, which specify the physical memory access privileges (read, write, execute) for each physical memory region. Lastly, timing channel mitigation aims to provide confidentiality by eliminating side-channel attacks.

Apart from a 1MB on-chip SRAM SPM, Shaheen connects to up to 512MB of off-chip low-power HyperRAMs~\cite{hyperram_low_pincount} directly on the main interconnect, through a low-power, low-cost 0.27$mm^2$ \SI{1.6}{\giga bp \second} fully-digital memory controller.
Relying on HyperRAMs instead of high-end LPDDR4/5 memories, typically integrated into embedded Linux-capable systems, frees Shaheen from expensive and proprietary memory controllers with large mixed-signal PHYs, while still exposing hundreds of MB to the host processor and matching the tight power, form factor and cost requirements of nano-UAVs.

The cluster integrates the so-called Flex-V cores.
The Flex-V core extends the RISC-V ISA with custom instructions for reduced-precision single instruction multiple data (SIMD) FP-based computation and byte and sub-byte mixed-precision QNN inference, achieving State-of-the-Art (SoA) software power and energy efficiency.
%
Thanks to the aggressive optimizations, the cluster achieves up to 22.5 GigaOperations per second (GOp/s) and \SI{90}{\giga Op/\second} on 8-bit and 2-bit integer kernels, enabling low-latency mixed-precision QNN-based autonomous navigation~\cite{dronet, cereda2023deep}.
Furthermore, the cluster achieves up to 4 GigaFloating-Point Operations per second (GFLOp/s) and \SI{7.9}{\giga FLOp/\second} on FP32 and FP16 kernels enabling DSP and online training of neural networks (NNs)\cite{amodei2016concrete, trainlib_git}.

To sum up, compared to the State-of-the-Art MCUs for nano-UAVs, Shaheen is the first one coupling:
\begin{itemize}
    \item an RV64 host with Hypervisor support and security features,
    \item a low-power memory controller exposing hundreds of MB at up to 1.6Gbps to the host core,
    \item a fully-programmable parallel RV32 cluster providing SoA software performance for IoT,
\end{itemize}
while keeping the overall power envelope within $200$mW.

The structure of this manuscript is as follows: in Section \ref{sec:related}, we will present an overview of the State-of-the-Art SoC for UAVs. Following this, Sections \ref{sec:arch} and \ref{sec:impl} will delve into an exhaustive discussion of Shaheen's architecture, its implementation, and the measurements obtained from the silicon prototype. Moving forward, Sections \ref{sec:swstack} and \ref{sec:bench} will address the software stack and provide a benchmark of the cluster's performance and energy efficiency on a relevant set of applications for nano-UAVs. In Section \ref{sec:comp_silicon}, we will compare Shaheen with similar silicon prototypes from both industry and academia. Finally, Section \ref{sec:concl} will summarize our results and offer insights into potential future research directions.

\section{Related work}\label{sec:related}

\begin{table}[t]
    \centering
    \caption{UAVs taxonomy by vehicle class-size \cite{dronet}.}
    \begin{tabular}{llll} \hline
          \multirow{3}{*}{\begin{tabular}[l]{@{}l@{}}Vehicle \\ class size\end{tabular}} &
          \multirow{3}{*}{\begin{tabular}[l]{@{}l@{}} Diameter : Weight \\ $[cm : kg]$\end{tabular}} &
          \multirow{3}{*}{\begin{tabular}[l]{@{}l@{}}Power Budget \\ $\frac{\text{Total[W]}}{\text{Compute[W]}}$\end{tabular}} &
          \multirow{3}{*}{\begin{tabular}[l]{@{}l@{}}Onboard\\ computer\end{tabular}}   \\ 
          &  &  & \\ 
          &  &  & \\ \hline
         \textit{standard}\cite{std_jetson} & $\sim$ 50 : $\geq$ 1       & $\geq$ 100  / 5-15          & Desktop/Emb.   \\
         \textit{micro}\cite{agilicious}    & $\sim$ 25 : $\sim$ 0.5  & $\sim$ 50 / 2.5-7.5      & Embedded  \\
         \textit{nano}\cite{dronet}         & $\sim$ 10 : $\sim$ 0.01 & $\sim$ 5  / 0.25-0.75  & MCU       \\
         \textit{pico}\cite{pico}           & $\sim$  2 : $\leq$ 0.001   & $\sim$ 0.1 / 5-15E-3 & ULP       \\ \hline
    \end{tabular}
    \label{tab:uav_taxonomy}
\end{table}

Table \ref{tab:uav_taxonomy} shows the four categories of UAV systems according to size, weight, power budget and onboard processing platform.
The latter two characteristics are tightly coupled, as only around $\sim 5-15\%$ of the power budget is allocated to computation\cite{dronet}.
Across all the categories of UAVs, autonomous navigation is achieved by the combination of two components: mission control and flight control.
Mission control is the high-level decisional part of the navigation algorithm, e.g., path planning~\cite{forsberg2017gpu}, optimization-based control~\cite{QUINTERO201428}, etc.
To carry out these types of tasks, SoA drones mostly rely on machine learning (ML) algorithms\cite{agilicious, dronet}.
Flight control, on the other hand, is the actuation of the output decisions of mission control: it collects data from the sensors to determine the vehicle's state and generates the control law, which manages the actuators~\cite{px4}.
Flight control is often based on cascade PID control~\cite{idrissi2022review}, especially in the case of nano-UAVs~\cite{budaciu,zekry}, and it is not as computationally intensive as mission control,  but it requires low-latency guarantees. 
As a consequence, also in the context of standard and micro-drones, flight control is usually carried out by simple MCUs with a predictable execution time like the STM32-H7\cite{stm32h7} integrated into the Pixhawk board~\cite{px4}.
Table~\ref{tab:related-socs} shows some mainstream SoCs successfully deployed on drones of standard, micro, and nano size.
For each SoC, it highlights the different computational capabilities and power envelope, as well as the specific tasks and UAV platforms they are suited for, detailed in the three sections below.
Sections~\ref{ssec:std-micro} and~\ref{ssec:nano} describe the state of the art of standards, micro, and nano UAVs SoCs.

\begin{table*}[t]
\centering
\caption{State-of-the-Art SoCs for UAVs.}
\label{tab:related-socs}
\begin{tabular}{|c|c|c|c|c|c|c|c|c|c|}
\hline
\multirow{3}{*}{\textbf{\begin{tabular}[c]{@{}c@{}}SoC \\ (Proxy price)\end{tabular}}} &
  \multirow{3}{*}{\textbf{\begin{tabular}[c]{@{}c@{}}Target \\ Platform\end{tabular}}} &
  \multirow{3}{*}{\textbf{Task}} &
  \multirow{3}{*}{\textbf{\begin{tabular}[c]{@{}c@{}}CPU : \\ Max freq \\{[}MHz{]}\end{tabular}}} &
  \multirow{3}{*}{\textbf{\begin{tabular}[c]{@{}c@{}}Supported \\ OS \end{tabular}}} &
  \multirow{3}{*}{\textbf{\begin{tabular}[c]{@{}c@{}}HW Virt. \\ support \end{tabular}}} &
  \multirow{3}{*}{\textbf{\begin{tabular}[c]{@{}c@{}}Parallel  \\ accelerator\end{tabular}}} &
  \multirow{3}{*}{\textbf{\begin{tabular}[c]{@{}c@{}}Accelerator \\ FLOp/s\end{tabular}}} &
  \multirow{3}{*}{\textbf{\begin{tabular}[c]{@{}c@{}}Power \\ envelope : \\ Technology \end{tabular}}} &
  \multirow{3}{*}{\textbf{\begin{tabular}[c]{@{}c@{}}Main \\ memory\end{tabular}}} \\
 &
   &
   &
   &
   &
   &
   &
   &
   &
   \\
    &
   &
   &
   &
   &
   &
   &
   &
   &
   \\ \hline
\multirow{3}{*}{\textit{\begin{tabular}[c]{@{}c@{}}NVIDIA Jetson \\ TX2 \cite{Jetson}\\ (100-150\$)\end{tabular}}} &
  \multirow{3}{*}{\begin{tabular}[c]{@{}c@{}}Standard/ \\ micro-size\end{tabular}} &
  \multirow{3}{*}{Mission} &
  \multirow{3}{*}{\begin{tabular}[c]{@{}c@{}}4×Cortex-A57 :\\ 2 GHz\end{tabular}} &
  \multirow{3}{*}{Linux} &
  \multirow{3}{*}{\checkmark} &
  \multirow{3}{*}{\begin{tabular}[c]{@{}c@{}}256x Pascal\\ CUDA\end{tabular}} &
  \multirow{3}{*}{\begin{tabular}[c]{@{}c@{}}1.33 \\ TFLOp/s\end{tabular}} &
  \multirow{3}{*}{\begin{tabular}[c]{@{}c@{}}7.5-15W :\\ 16nm\end{tabular}} &
  \multirow{3}{*}{\begin{tabular}[c]{@{}c@{}}8GB \\ LPDDR4\end{tabular}} \\
 &
   &
   &
   &
   &
   &
   &
   &
   &
   \\
    &
   &
   &
   &
   &
   &
   &
   &
   &
   \\
   \hline
\multirow{3}{*}{\textit{\begin{tabular}[c]{@{}c@{}}Intel Atom \\ x7-E3950\cite{intel} \\ (50\$)\end{tabular}}} &
  \multirow{3}{*}{\begin{tabular}[c]{@{}c@{}}Micro-size\end{tabular}} &
  \multirow{3}{*}{Mission} &
  \multirow{3}{*}{\begin{tabular}[c]{@{}c@{}}4xIntel Atom : \\ 2GHz\end{tabular}} &
  \multirow{3}{*}{Linux} &
  \multirow{3}{*}{\xmark} &
  \multirow{3}{*}{\begin{tabular}[c]{@{}c@{}}GPU Intel HD \\ 505\end{tabular}} &
  \multirow{3}{*}{\begin{tabular}[c]{@{}c@{}}230\\GFLOp/s\end{tabular}} &
  \multirow{3}{*}{\begin{tabular}[c]{@{}c@{}}$\sim$10W :\\ 14nm\end{tabular}} &
  \multirow{3}{*}{\begin{tabular}[c]{@{}c@{}}8GB \\ LPDDR3\end{tabular}} \\
 &
   &
   &
   &
   &
   &
   &
   &
   &
   \\ 
 &
   &
   &
   &
   &
   &
   &
   &
   & \\\hline
\multirow{3}{*}{\textit{\begin{tabular}[c]{@{}c@{}}Allwinnner \\ H3 \cite{agilicious} \\ (\textless{}10\$)\end{tabular}}} &
  \multirow{3}{*}{\begin{tabular}[c]{@{}c@{}}Micro-size\end{tabular}} &
  \multirow{3}{*}{Mission} &
  \multirow{3}{*}{\begin{tabular}[c]{@{}c@{}}4xCortex-A7 : \\ 1.2GHz\end{tabular}} &
  \multirow{3}{*}{Linux} &
  \multirow{3}{*}{\checkmark} &
  \multirow{3}{*}{\begin{tabular}[c]{@{}c@{}}GPU Mali-400 \\ MP2\end{tabular}} &
  \multirow{3}{*}{\begin{tabular}[c]{@{}c@{}}10\\GFLOp/s\end{tabular}} &
  \multirow{3}{*}{\begin{tabular}[c]{@{}c@{}}\textgreater{}1W :\\ 40nm\end{tabular}} &
  \multirow{3}{*}{\begin{tabular}[c]{@{}c@{}}512MB \\ LPDDR3\end{tabular}} \\
 &
   &
   &
   &
   &
   &
   &
   &
   &
   \\
    &
   &
   &
   &
   &
   &
   &
   &
   &
   \\
   \hline
\multirow{3}{*}{\begin{tabular}[c]{@{}c@{}} \textit{STM32-H7} \cite{stm32h7} \\ (15\$)\end{tabular}} &
  \multirow{3}{*}{\begin{tabular}[c]{@{}c@{}}Standard/ \\ micro-size \end{tabular}} &
  \multirow{3}{*}{Flight} &
  \multirow{3}{*}{\begin{tabular}[c]{@{}c@{}}Cortex M7 + \\ Cortex M4 : \\ 480 - 240\end{tabular}} &
  \multirow{3}{*}{Linux} &
  \multirow{3}{*}{\xmark} &
  \multirow{3}{*}{-} &
  \multirow{3}{*}{\begin{tabular}[c]{@{}c@{}}\textless{}500\\MFLOp/s\end{tabular}} &
  \multirow{3}{*}{\begin{tabular}[c]{@{}c@{}}\textless 200mW : \\ 40nm\end{tabular}} &
  \multirow{3}{*}{\begin{tabular}[c]{@{}c@{}}512kB \\ SRAM\end{tabular}} \\
 &
   &
   &
   &
   &
   &
   &
   &
   &
   \\
 &
   &
   &
   &
   &
   &
   &
   &
   &
   \\ \hline
\multirow{2}{*}{\begin{tabular}[c]{@{}c@{}} \textit{STM32-F4} \cite{stm32f4} \\ (15\$)\end{tabular}} &
  \multirow{2}{*}{Nano-size} &
  \multirow{2}{*}{Both} &
  \multirow{2}{*}{\begin{tabular}[c]{@{}c@{}}Cortex M4 : \\ 180\end{tabular}} &
  \multirow{2}{*}{RTOS} &
  \multirow{2}{*}{\xmark} &
  \multirow{2}{*}{-} &
  \multirow{2}{*}{-} &
  \multirow{2}{*}{\begin{tabular}[c]{@{}c@{}}\textless 200mW : \\ 90nm\end{tabular}} &
  \multirow{2}{*}{\begin{tabular}[c]{@{}c@{}}1MB \\ SRAM\end{tabular}} \\
    &
   &
   &
   &
   &
   &
   &
   &
   &
   \\
   \hline
\multirow{2}{*}{\begin{tabular}[c]{@{}c@{}}\textit{GAP8\cite{gap8}} \\ (40\$)\end{tabular}} &
  \multirow{2}{*}{Nano-size} &
  \multirow{2}{*}{Both} &
  \multirow{2}{*}{Ri5cy : 250} &
  \multirow{2}{*}{RTOS} &
  \multirow{2}{*}{\xmark} &
  \multirow{2}{*}{8xRi5cy} &
  \multirow{2}{*}{-} &
  \multirow{2}{*}{\begin{tabular}[c]{@{}c@{}}\textless 100mW : \\ 55nm\end{tabular}} &
  \multirow{2}{*}{\begin{tabular}[c]{@{}c@{}}1.5MB \\ SRAM\end{tabular}} \\
 &
   &
   &
   &
   &
   &
   &
   &
   &
   \\ \hline
\multirow{2}{*}{\begin{tabular}[c]{@{}c@{}}\textit{GAP9\cite{gap8}} \\ (40\$)\end{tabular}} &
  \multirow{2}{*}{Nano-size} &
  \multirow{2}{*}{Both} &
  \multirow{2}{*}{Ri5cy : 450} &
  \multirow{2}{*}{RTOS} &
  \multirow{2}{*}{\xmark} &
  \multirow{2}{*}{9xRi5cy} &
  \multirow{2}{*}{\begin{tabular}[c]{@{}c@{}}\textless{}3GFLOp/s\\(FP32)\end{tabular}} &
  \multirow{2}{*}{\begin{tabular}[c]{@{}c@{}}\textless 50mW : \\ 22nm\end{tabular}} &
  \multirow{2}{*}{\begin{tabular}[c]{@{}c@{}}1.5MB \\ SRAM\end{tabular}} \\
 &
   &
   &
   &
   &
   &
   &
   &
   &
   \\ \hline
\multirow{2}{*}{\begin{tabular}[c]{@{}c@{}}\textit{Kraken\cite{kraken}} \\ (\textit{Prototype})\end{tabular}} &
  \multirow{2}{*}{Nano-size} &
  \multirow{2}{*}{Both} &
  \multirow{2}{*}{\begin{tabular}[c]{@{}c@{}}Ri5cy-NN : \\ 350\end{tabular}} &
  \multirow{2}{*}{RTOS} &
  \multirow{2}{*}{\xmark} &
  \multirow{2}{*}{8xRi5cy-NN} &
  \multirow{2}{*}{\begin{tabular}[c]{@{}c@{}}\textless{}3GFLOp/s\\(FP32)\end{tabular}} &
  \multirow{2}{*}{\begin{tabular}[c]{@{}c@{}}\textless 300mW : \\ 22nm\end{tabular}} &
  \multirow{2}{*}{\begin{tabular}[c]{@{}c@{}}1MB \\ SRAM\end{tabular}} \\
 &
   &
   &
   &
   &
   &
   &
   &
   &
   \\ \hline
\multirow{2}{*}{\begin{tabular}[c]{@{}c@{}}\textit{This Work} \\ (\textit{Prototype})\end{tabular}} &
  \multirow{2}{*}{Nano-size} &
  \multirow{2}{*}{Both} &
  \multirow{2}{*}{RV-64 : 600} &
  \multirow{2}{*}{\begin{tabular}[c]{@{}c@{}}Linux\\+RTOS\end{tabular}} &
  \multirow{2}{*}{\checkmark} &
  \multirow{2}{*}{8xFlex-V} &
  \multirow{2}{*}{\begin{tabular}[c]{@{}c@{}}7.9GFLOp/s\\(FP16)\end{tabular}} & 
  \multirow{2}{*}{\begin{tabular}[c]{@{}c@{}}\textless 200mW : \\ 22nm\end{tabular}} &
  \multirow{2}{*}{\begin{tabular}[c]{@{}c@{}}8-512MB \\ HyperRAM\end{tabular}} \\
 &
   &
   &
   &
   &
   &
   &
   &
   &
   \\ \hline
\end{tabular}
\end{table*}

\subsection{SoCs for Standard and Micro-sized UAVs}\label{ssec:std-micro}

As table \ref{tab:uav_taxonomy} shows, micro-size drones integrate embedded computers, while standard-sized drones can even accommodate desktop processors.
Nevertheless, embedded processors can nowadays deliver performance in the order of hundreds of TOp/s and hundreds of GFLOp/s, which has proven to be sufficient to support the full flight stack for mission control, both for micro\cite{agilicious} and standard-size UAVs\cite{std_jetson}.
%

Embedded computers integrate high-end SoCs with application-class cores, supporting virtualization and various privilege levels (and hence full-fledged OSs), embedded GPUs, and GBytes of high-performance off-chip LPDDR/DDR4/5 memories, connected through expensive, large and power-hungry mixed-signal DDR controllers, all within a power envelope of few watts~\cite{Jetson,esp_tapeout,2dla-isscc,8_vpus}.

The NVIDIA Jetson TX2 is claimed to be \textit{"the fastest, most power-efficient embedded artificial intelligence (AI) computing device"} by NVIDIA\cite{Jetson} and it is the board of choice for the Agilicious drone\cite{agilicious}.
It features a  Quad-core Cortex A57 running up to 2GHz and a Pascal CUDA GPU, which can deliver up to 1.33TFLOp/s resulting in an overall power consumption of more than 7.5 W.
The Intel Atom x7 is the heart of the Intel UpBoard platforms, as big as a credit card.
It features 4 Intel Atom processors running up to 2GHz and an Intel HD 505 GPU delivering up to 230GFLOp/s and roughly consuming 10W.
Another compute hardware platform commonly used on autonomous UAVs is the NanoPi Neo Air, which integrates an Allwinner H7 SoC with a quad-core CortexA7 and a Mali-400 MP2 GPU, delivering up to 10GFLOp/s.
All these SoCs offer a mainstream Ubuntu-ready software stack and virtualization capabilities and can handle very sophisticated and complex applications.
However, due to their power envelope, size, and the necessity for high-end off-chip memories, these SoCs can only be integrated into standard and micro-UAVs.

Naturally, Shaheen can not compete with these architectures in terms of performance, but our approach borrows the best of their characteristic while targeting a much smaller power envelope.
Firstly, to mimic high-end SoCs with their heterogeneous GPU-based architecture, Shaheen integrates an RV32-based parallel programmable cluster along with an RV64 CPU.
Secondly, it exposes a significant amount of off-chip main memory to the CPU.
However, instead of high-performance DRAMs (LPDDR3/4/5) that are connected through large, proprietary and expensive mixed-signal PHYs with a high pin count ($>$30), Shaheen leverages HyperRAMs, which are fully-digital low-power small-area DRAMs with less than 14 pins and feasible to be deployed on nano-UAVs.
A similar approach is adopted in Cheshire\cite{cheshire}, which is not optimized for nano-UAV applications. Cheshire revolves around CVA6 as Shaheen and exposes up to 1GB of Reduced Pin Count (RPC) DRAM memory, which uses a minimum number of signals to deliver DDR3-level in-system bandwidth at the cost of 22 switching signals for a 16-bit wide data bus\cite{etron,cheshire}. While RPC and the related controller offer higher bandwidth than HyperBUS, the RPC protocol is more convoluted, leading to higher design complexity and a bigger area, mostly due to the four 8kB buffers\cite{cheshire}. More importantly, Cheshire's CVA6 does not feature hardware virtualization support and micro-architectural extensions for timing channel mitigations. Lastly, while being easily extensible through the AXI4 interface, Cheshire's silicon prototype does not integrate a parallel accelerator, heavily limiting the offered performance.
To sum up, Shaheen is the first silicon implementation of a heterogeneous MCU coupling an RV64 core with a cluster of eight RV32 cores and tens of MB of main memory.

\subsection{SoCs for Nano-UAVs}\label{ssec:nano}

A state-of-the-art MCU for nano-UAVs platforms is the STM32-F4\cite{stm32f4}.
The STM32-F4 is the computational unit of the Crazyflie\cite{crazyflie} platform, integrating a Cortex-M4 core and 192kB of on-chip SRAM with 180MHz of maximum operating frequency.
Its low performance and small memory capacity limit the autonomous navigation capabilities of the nano-drone when compared to embedded computers. 
To this extent, two kinds of approaches have been proposed: minimization of the workload \cite{neural-swarm} or offloading of the mission control computation to an external base station\cite{fuzzy}, limiting the MCU to flight control.
The latter approach presents severe drawbacks, in the first instance, it introduces network-dependent latency, limiting the maximum distance from the workstation to a few tens of meters. 
Also, the data transmitted are subject to noise on the transmission channel, limiting reliability, and eavesdropping on confidential data\cite{SOK}.

To offer enhanced computational capabilities within a small power budget, recent works also propose SoCs featuring hardwired ASIC accelerators designed for specific UAV applications, like, for example, motion-control~\cite{motion-control-isscc}, visual-inertial odometry (VIO)~\cite{navion}, simultaneous-localization-and-mapping (SLAM)\cite{slam}, or QNN inference \cite{gap8,kraken}.
These accelerators achieve impressive energy efficiency, in the order of hundreds of TOp/s/W, by carefully mapping the target algorithm to the hardware.
For example, many accelerators exploit the inherent parallelism of the target application, such as using a systolic array for motion control\cite{motion-control-isscc}.
Another common approach exploits reduced-precision arithmetic, as in SLAM\cite{slam} and VIO\cite{navion}, to reduce the memory footprint and the datapath size.
Exploiting both parallelism and reduced-precision computation is also a well-established technique to accelerate QNN inference and training, due to the nature of such algorithm.
For example, accelerators like the NE16 in GAP9\cite{gap8}, the HWCE in GAP8\cite{gap8}, and the ternary weight neural-network (so-called CUTIE) accelerator in Kraken\cite{kraken}, able to reach peaks of 11.6 TMAC/s, have been proposed to speed up QNN inference.
However, due to poor flexibility and programmability, these accelerators have to anyway rely on general-purpose CPUs to achieve end-to-end flight.
Furthermore, the high area cost per device makes them hard to adopt as they risk becoming obsolete due to the rapid evolution of the target nano-UAVs applications. 

To overcome these limitations, recent MCUs integrate parallel fully-programmable and flexible accelerators, that have successfully proved to enable autonomous navigation\cite{dronet,cereda2023deep}.
Namely, GAP8, GAP9\cite{gap8} and Kraken\cite{kraken} are MCUs with enhanced computational capabilities, based on parallel programmable accelerators.
GAP8 and GAP9 are commercial products by GreenWaves Technologies compliant with the so-called Crazyflie-AIdeck\cite{crazyflie-ai} board, which is meant as a companion of the Crazyflie to offload the mission control tasks\cite{dronet}.
GAP8 embeds the so-called Ri5cy\cite{gap8} core as host CPU and 1.5MB of on-chip SRAM memory, accompanied by a parallel programmable cluster of other eight Ri5cy cores delivering up to 150 GOP/s on 8-bit data.
Ri5cy is a 4 pipeline-stages core compliant with the so-called XpulpV2 ISA, a custom RISC-V ISA based on RV32 with extensions for DSP and ML applications, with support for 16/8-bit SIMD operations and hardware loops.
GAP9 is an improved version of the GAP8 processor.
It is fabricated in a more advanced node than GAP8, halving the power envelope and it features as well 2MB of non-volatile SVM memory.
Also, differently from GAP8, GAP9's cluster includes 4 FPUs with FP16/32 support.
Lastly, Kraken \cite{kraken} is a research prototype based on the same heterogeneous architecture of GAP8 and GAP9, i.e., an RV32 CPU along with an eight RV32 core cluster, which delivers up to 90 GOp/s on 2-bit data.
Kraken's RV32 cores are a more advanced version of Ri5cy, i.e., the Ri5cyNN cores \cite{kraken} with support of sub-byte SIMD operations and fused \textit{Mac\&Load} instructions, which enable the concurrent execution of SIMD dot-product and memory accesses, increasing the computation efficiency up to $94\%$.
Kraken embeds 1.5MB of on-chip SRAM memory and the CUTIE accelerator, able to achieve up to roughly 90k Ternary-MACs per cycle.
Furthermore, it provides an event-based camera, tightly coupled with a Spiking Neural Engine accelerator. 
When compared to traditional cameras, event-based cameras offer high temporal resolution (in the order of $\mu$s), very high dynamic range (140 dB vs. 60 dB), low power consumption, and high pixel bandwidth (on the order of kHz) resulting in reduced motion blur\cite{event_based}.

Shaheen's approach leverages the best from these advanced AI IoT SoCs, integrating its own fully-programmable parallel 8-core RV32 cluster accelerator.
Shaheen's RV32 cores stem from the Ri5cyNN cores and are further enhanced with mixed-precision support to eliminate the massive software overhead necessary for packing and unpacking data when executing mixed-precision sub-byte kernels, providing up to 8.5x speed-up over Kraken and less than $5.6\%$ extra area resource over the baseline core without extensions.
In addition, Shaheen addresses a major limitation of SoA MCUs: the software stack based on lightweight RTOSs or simple bare-metal runtimes.
Programming applications on these stacks is hard, owing to (i) the lack of virtualization capabilities of the host CPUs and (ii) the small amount of memory directly accessible through loads and stores, which limits the maximum software memory footprint.
In the context of MCUs, memory resources coincide with on-chip SRAMs and off-chip DRAMs.
The first ones provide high bandwidth but are limited to a few hundred kB, due to the area and power constraints~\cite{hyperram_low_pincount}.
The latter ones offer one order of magnitude more capacity but are much slower and are typically accessed only through explicit input-output copy functions.
Thus, beyond SoA, to support a richer software stack while offering the advanced computing performance and energy efficiency of the RV32-based cluster, Shaheen integrates an RV64 core with advanced virtualization and security features, along with up to 512MB of main memory.
This enables the secure coexistence of rich and mature general-purpose OS and bare-metal RTOS on the same platform and eases porting of feature-rich software stacks for robotics, such as ROS\cite{ROS}.


\begin{figure}[t]
    \centering
    \includegraphics[width=0.95\linewidth]{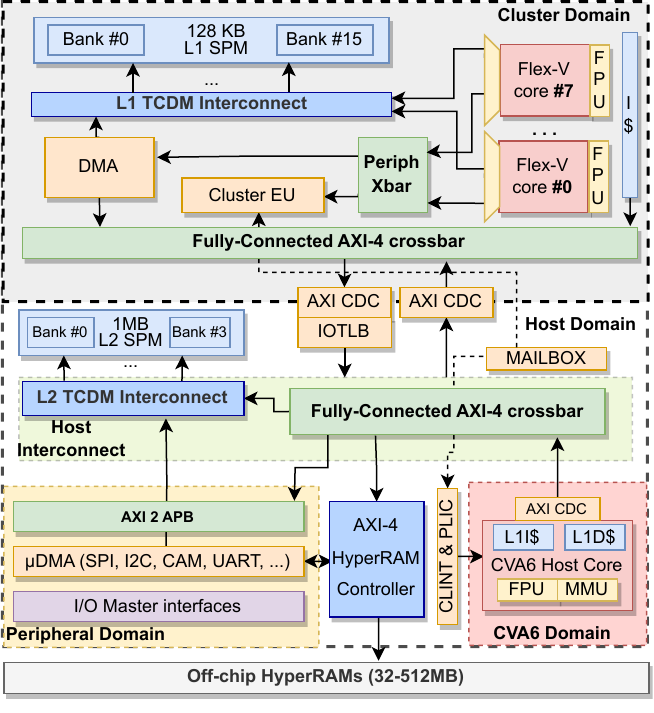}
    \caption{Shaheen architecture block diagram.}
    \label{fig:block}
\end{figure}

\section{Shaheen Architecure}\label{sec:arch}

Shaheen consists of 4 clock domains, as illustrated in Fig. \ref{fig:block}: (i) the CVA6 domain, where the host core is; (ii) the host domain, including the main interconnect and 4 256kB interleaved SRAM banks; (iii) the cluster domain, served by 16 16kB interleaved SRAM banks and 8 specialized RV32 cores; and (iv) the peripheral domain.

\subsection{CVA6 host core}

CVA6 \cite{cva6_git} is the heart of Shaheen.
It is an open-source 6-stages, single-issue, in-order, 64-bit Linux-capable RISC-V core, supporting the RV64GC ISA variant, SV39 virtual memory with a dedicated Memory Management Unit (MMU), three levels of privilege (Machine, Supervisor, User), and PMP\cite{pie}.
In the context of this work, the baseline version of CVA6 has been enhanced with 2 extra features to provide high-assurance isolation between the different applications co-existing on the core: 
\begin{enumerate}
    \item hardware support for virtualization compliant with the ratified 1.0 version of the RISC-V Hypervisor specification\cite{bruno_h}.
    \item temporal fence instruction, namely \textit{fence.t}~\cite{nils_fence}, which flushes $\mu$architectural state and enables the OS to close covert channels with a low increase in context switch costs and negligible hardware overhead.
\end{enumerate}
Such new features are relevant to many UAV applications such as the co-existence of full-fledged and real-time OSes (both custom and legacy), as well as isolation for safety and security reasons.

\subsubsection{H extension}

\begin{figure}[t]
    \centering
    \includegraphics[width=\linewidth]{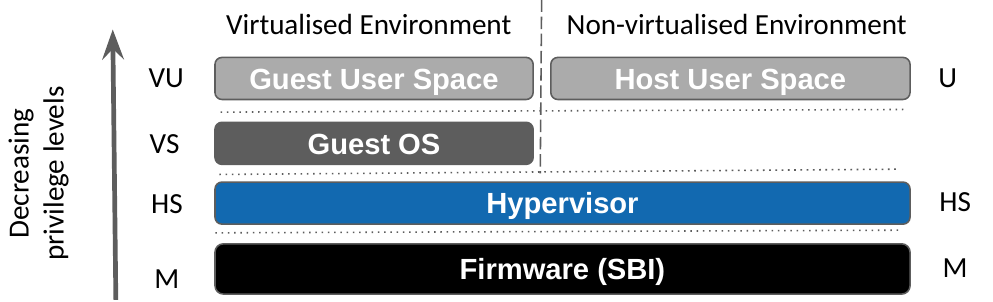}
    \caption{RISC-V privilege levels.}
    \label{fig:priv_stack}
\end{figure}

Tab. \ref{tab:priv_modes} shows the different privilege modes when implementing the Hypervisor extension and Fig.~\ref{fig:priv_stack} the resulting software stack.
The \textit{nominal privilege} modes are \textit{machine} (M), \textit{supervisor} (S), and \textit{user} (U). 
The Hypervisor extension adds the \textit{virtualization mode} (V), indicating whether the hart is currently executing in a guest (V=1) or not (V=0).
When V=0, the S-mode is modified into the \textit{hypervisor-extended supervisor} (HS) mode, well suited to host both type-1 and type-2 hypervisors.
Other than in the HS-mode, when V=0, the hart can either be in M-mode or in U-mode atop an OS running in HS-mode.
When V=1, two new privilege levels are added, namely the \textit{virtual supervisor} (VS) mode and the \textit{virtual user} (VU) mode.
Also, the hypervisor extension defines a second stage of translation (the so-called "G-stage") to virtualize the guest memory by translating guest-physical addresses (GPA) into host-physical addresses (HPA). 

\begin{table}[t]
    \centering
    \begin{tabular}{|c|c|c|c|} \hline
        Virtualization & Nominal   & \multirow{2}{*}{Name}        & 2-stage     \\
        Mode (V)       & Privilege &                              & translation \\ \hline
        0              & U         & User (U) mode                & Off         \\ \hline
        0              & S         & Hypervisor-extended          & Off         \\ 
                       &           & Supervisor (HS) mode         &             \\ \hline
        0              & M         & Machine (M) mode             & Off         \\ \hline
        1              & U         & Virtual User (VU) mode       & On          \\ \hline
        1              & S         & Virtual Supervisor (VS) mode & On          \\ \hline
    \end{tabular}
    \caption{RISC-V ISA privilege modes with the Hypervisor extension.}
    \label{tab:priv_modes}
\end{table}

To enable these new execution modes, the Control Status Register (CSR) and Decode modules have been modified.
The CSR module was extended to implement the first three building blocks that comprise the hardware virtualization logic, specifically: (i) access logic and permission checks for VS-mode and HS-mode CSRs, (ii) delegation and triggering of exceptions and interrupts, and (iii) handling of trap entry and exit.
The Decode module underwent changes to enable the decoding of hypervisor instructions (such as hypervisor load/store instructions and memory-management fence instructions), as well as the execution of all VS-mode-related instructions and the triggering of access exceptions. 

%
%
The MMU's page table walker (PTW) and translation lookaside buffer (TLB) have been modified to support the second stage of translation.
The PTW features a new control state to monitor the current stage of translation and facilitate the switching of contexts between VS-Stage and G-Stage translations.
Finally, the TLB entries have been extended to store both VS-Stage and G-Stage Page Table Entries (PTEs), as well as their corresponding permissions and virtual machine identifiers.
Overall, all these modifications account for less than $6\%$ extra area and hardware while enabling the safe co-existence of a full-fledge guest-OS (executed in VS-mode) together with a bare-metal RTOS (executed in U-mode).

\subsubsection{\textit{fence.t}}

\begin{figure}[t]
    \centering
    \includegraphics[width=\linewidth]{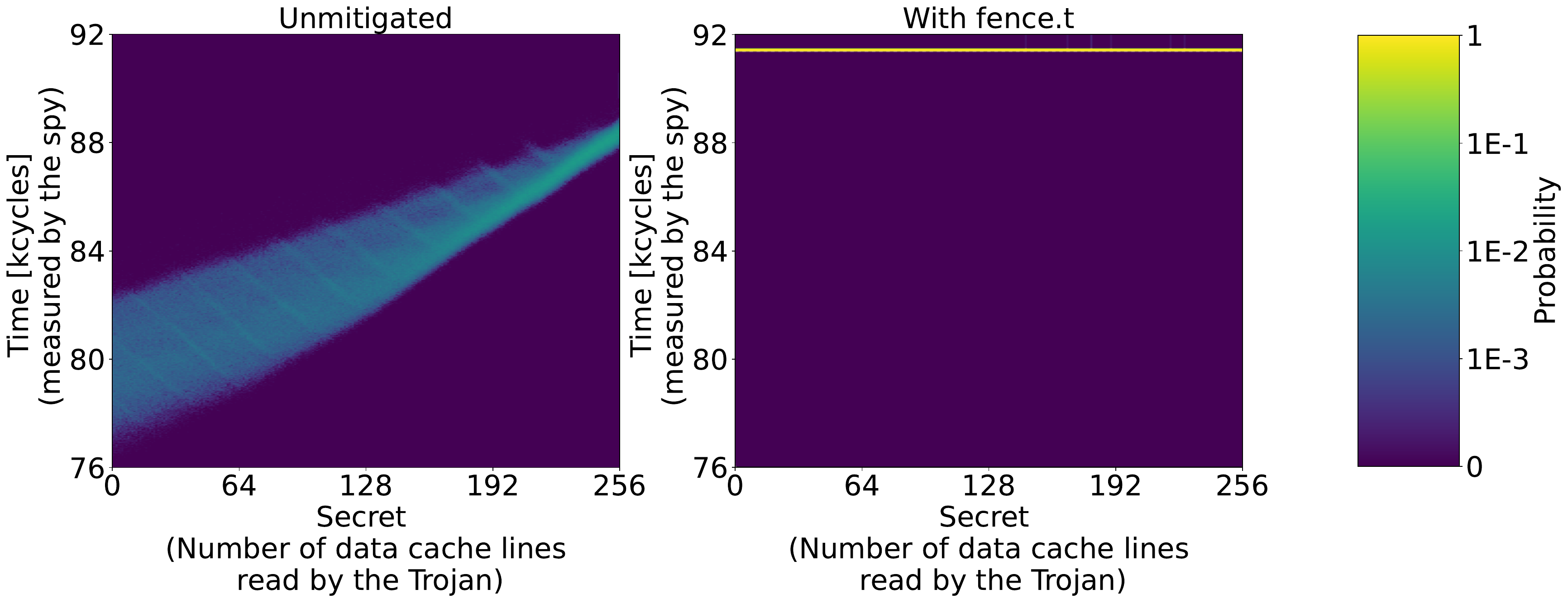}
    \caption{Channel matrices on the \textit{CHANNEL BENCH} test.}
    \label{fig:fence_t}
\end{figure}

The CVA6 core also implements the \textit{fence.t} instruction ~\cite{nils_fence}.
This instruction is added to CVA6's ISA to prevent \textit{timing channels}: exploitable hardware resources holding state depending on the execution history (caches, TLBs, branch predictors, and prefetchers) can leak information if not properly reset during a context switch.
Timing channels can be exposed by \textit{prime-and-probe} attacks~\cite{nils_fence}.
In these kinds of attacks, the spy first brings the target hardware resource into a known state (\textit{prime}).
In the following time slice, the OS switches to an application containing a Trojan, which accesses a subset of the hardware resource to encode a secret. %
Finally, when the execution switches back to the spy, it again probes (\textit{probe}) the whole buffer and observes an execution time $t$, correlated with the encoded secret. 
For data caches, the spy traverses a large buffer of $n$ lines so that the Trojan can then transmit a secret $s \leq n$ by touching $s$ lines: in the last time slice, the spy decodes $s$ from the measured execution time $t$.

In this context, the \textit{fence.t} extends the control that the OS, or the Hypervisor, has over the hardware. 
Namely, it provides the capability of clearing vulnerable microarchitectural states to enable a history-independent context-switch latency by flushing the caches and the TLB and resetting the internal FSMs of the core.
The \textit{fence.t} has been validated against prime-and-probe attacks from the \textit{MASTIK} toolkit~\cite{yarom2016mastik,nils_fence}. 
These attacks are implemented within Ge's \textit{CHANNEL BENCH}~\cite{ge2019principled,channel_bench_git} suite, which provides a minimal OS and data collection infrastructure, running on an experimental version of seL4 supporting timing protection.
To visualize the correlation between $s$ and $t$, we use channel matrices. A channel matrix represents the conditional probability of getting an execution time $t$, having an input secret $s$.
In Fig.~\ref{fig:fence_t}, we represent the channel matrix as heatmaps: $s$ (the secret encoded by the Trojan by touching $s \leq n$ data cache lines) varies horizontally, and $t$ (the execution time measured by the spy) varies vertically, bright colours indicate a high probability and dark colours indicate a low probability of measuring such $t$, given a certain $s$.

Fig.~\ref{fig:fence_t} shows the channel matrices on the \textit{CHANNEL BENCH} test for CVA6's write-through L1 data cache. On the left it is shown the matrix when not using the \textit{fence.t}: the correlation between the Trojan's secret and the spy's probe time indicates a covert channel. On the right, when using the \textit{fence.t}, there is no correlation.
With less than 320 additional clock cycles to the context-switch latency (insignificant at typical switch rates of 1 kHz), the \textit{fence.t} requires a low implementation effort and negligible hardware costs.

\subsection{Host \& Peripheral Domain}

The host domain leverages the popular AXI4 protocol \cite{ARMAXI} for the main interconnect. Namely, it includes a 64-bit AXI4 crossbar delivering up to 32Gbps on each AXI4 port, respectively on read and write channels.
It also includes 4 256kB SRAM banks, composing a 1MB L2 ScratchPad Memory (L2SPM) delivering up to 64Gbps, either for writing or reading. 
The L2SPM is meant to (i) store data to be shared with off-chip peripherals, (ii) store the cluster code, (iii) for fast communication between CVA6 and the cluster, and, more in general, (iv) for low-latency ($<$10 clock cycles) and predictable accesses.

To enable independent data transfer from peripherals to the SoC, Shaheen includes in the peripheral domain the so-called \textit{"$\mu$DMA subsystem"} which is a controller intended to autonomously serve a set of I/O interfaces popular in critical applications.
Such interfaces include for instance HyperBUS, I2C, (Q)SPI, CPI, SDIO, UART, CAN, PWM, and I2S. The $\mu$DMA exports two ports, one for receiving and one for sending data, to read/write data from/to the L2SPM SRAM memory to/from the off-chip peripherals\cite{gap8}.
Shaheen also features an open-source Linux-compliant Ethernet IP, to be fully compliant with the Pixhawk standard \cite{px4}, popular open-source hardware specifications and guidelines for drone systems development.

\subsubsection{HyperRAM memory controller} 
\begin{figure}[t]
    \centering
    \includegraphics[width=\linewidth]{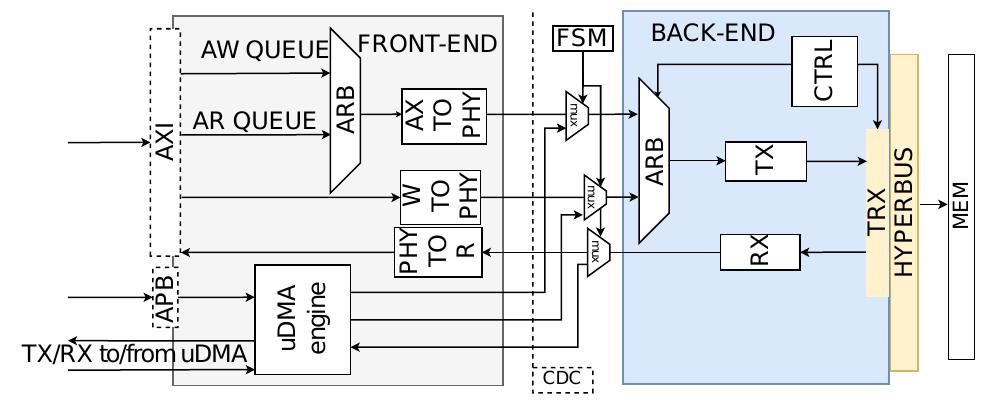}
    \caption{HyperRAM memory controller architecture.}
    \label{fig:hyper}
\end{figure}

Fig. \ref{fig:hyper} depicts Shaheen's HyperRAM controller, which provides a configuration APB port and an AXI4 subordinate port.
It connects the SoC with off-chip HyperRAMs, compliant with the HyperBUS protocol, which is a fully digital protocol counting $11+n$ pins: $3$ control pins, $n$ Chip Select (CS), and $8$ Double-Data-Rate pins used both for commands and data~\cite{hyperram_low_pincount}.
Depending on the off-chip memory models, the controller exposes between 32MB and 512MB to the interconnect, and it provides up to 1.6 Gbps.
HyperRAMs are the main memory of choice for Shaheen because, differently from high-end DDR DRAM memories, they target a much lower power consumption and silicon footprint while guaranteeing enough bandwidth for advanced AI IoT applications and capacity to boot embedded SPM Linux~\cite{hyperram_low_pincount}.

There are two distinct modules within the HyperRAM controller, i.e., the PHY controller (back-end) and the front-end, operating in different frequency domains.
The front-end module consists of an AXI4-to-PHY converter and a specialized $\mu$DMA engine channel accessible through APB to execute software-programmed DMA transfers.
The AXI4 and $\mu$DMA transactions are multiplexed towards the PHY, which translates the incoming data packets into HyperRAM transactions and vice versa.
The AXI4 front-end enqueues the AXI4 transactions individually and lets through only one read or one write at a time and converts it into a request for the PHY.
At this point, the back-end translates the request into a command for the HyperRAMs and issues it over the HyperBUS.
Following, in the case of a write, the W channel transactions get converted into multiple PHY data packets.
For reads, the PHY back-end sends data packets to a converter that then populates the R channel.
The $\mu$DMA engine directly connects the L2SPM and the back-end and can generate both 1D and 2D burst transactions.
These features are highly valuable for the efficient execution of ML algorithms on the cluster, as it is achieved through explicit orchestration of the data movement between the off-chip memory, the L2SPM and the L1SPM\cite{dory}.

To double the bandwidth and the capacity, Shaheen's back-end module controls 2 HyperBUS interfaces in parallel, and it controls 2 memories on each HyperBUS, with 2 dedicated CS.
Each memory is seen as a memory block of 16 bits width and $N$ rows, programmable at runtime according to the onboard memories available.
The pair of memories on the same CS of the two different buses are mapped as interleaved, hence occupying the first $2 \cdot 2 \cdot N$ Bytes.
The other pair of memory is placed contiguously on top.

\subsection{Parallel Programmable Cluster}

While the host core supports advanced virtualization, security and isolation features, it is not optimized for number crunching: when running computation-intensive kernels is needed, it invokes the cluster.
The cluster domain is a programmable parallel accelerator connected to the main host interconnect through a controller and a subordinate AXI4 port.
The cluster is composed of 8 70kGE 4-pipeline stages RV32 cores, optimized for general-purpose DSP and ML applications, described below.
The cores share 16 16kB interleaved SRAM banks, composing a 256kB L1 SPM, accessible through a single clock cycle latency logarithmic interconnect, providing up to 256 Gbps at 500MHz.
A hierarchical instruction cache, composed of 8 512 Bytes private caches and a 4kB of 2-cycle latency shared cache, assists the cores.
It is implemented with latch-based SCM to improve energy efficiency over energy-expensive SRAM cuts.
The cluster also includes a DMA with one 64-bit AXI4 port and 4 32-bit ports towards the L1SPM for high-bandwidth, low-latency transactions to/from the L1 SPM.
Leveraging explicit memory DMA transfers and scratchpad memories, double-buffering and custom ISA extensions, the cluster avoids the hardware overhead of expensive data caches while maximizing the utilization of memory and computing resources~\cite{dory}.

\subsubsection{Flex-V cores}

The 8 RV32 cores are the so-called Flex-V cores~\cite{flex_v}.
Each core has a dedicated FPU unit supporting FP32, FP16 and bfloat16 types, supporting SIMD instructions on lower precision data.
Also, all the cores share a single Floating Point division and square root operations unit (DIV/SQRT).
The Flex-V core is an aggressively optimized version of the Ri5cy core\cite{ri5cy_git}, which support the XPulpV2 ISA extension, considered as the baseline.
The Ri5cy core already provides custom instructions to accelerate the execution of ML and DSP workloads, namely, it supports post-increment LD/ST, hardware loops, and SIMD instructions down to 8-bit precision.
%

To enhance the performance of sub-byte uniform linear kernels, the XpulpNN ISA has been proposed~\cite{kraken}, which extends XpulpV2 ISA with 4- and 2-bit SIMD operations.
Additionally, it introduces fused \textit{Mac\&Load} instructions enabling simultaneous execution of SIMD dot-product operations alongside memory accesses, almost doubling the computation efficiency.
More precisely, the \textit{fused Mac\&Load (mlsdotp)} instruction combines a SIMD \textit{dot-product}-like operation with a load operation performed during the writeback stage.
Doing so enables replacing the non-stationary data in a register to directly feed the next \textit{Mac\&Load} instruction with it.
To decouple and simplify the \textit{Mac\&Load} execution, the XPulpNN core integrates six additional 32-bit registers, forming the so-called Neural Network Register File (NN-RF), enabling the Load operations (of weights and activations) during the \textit{Mac\&Load} write-back stage, which could not be performed otherwise on the general purpose register file (GP-RF). 
However, when dealing with mixed-precision inputs, the performance of XpulpNN degrades significantly because of the substantial software overhead required for packing and unpacking data.

\begin{figure}[t]
    \centering
    \includegraphics[width=0.9\linewidth]{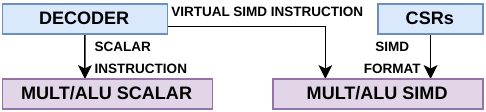}
    \caption{Instruction decoding during the status-based execution. }
    \label{fig:status-based}
\end{figure}

\begin{figure}[t]
    \centering
    \includegraphics[width=0.9\linewidth]{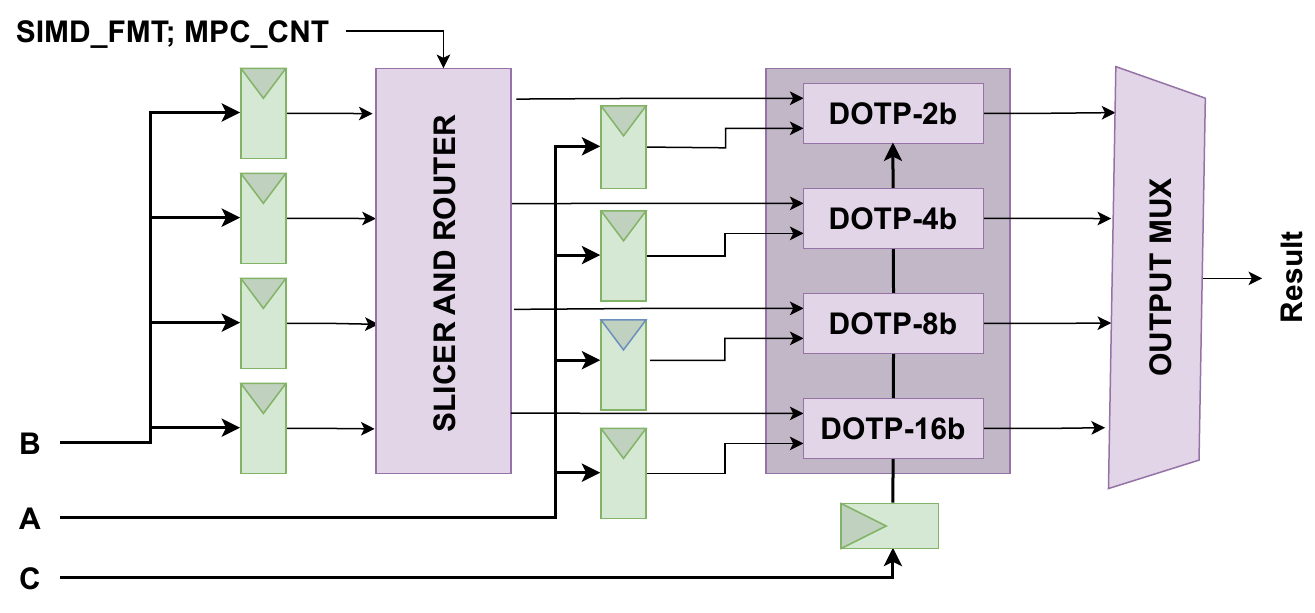}
    \caption{Dotp unit Datapath.}
    \label{fig:dotp-dp}
\end{figure}

\begin{figure}[t]
    \centering
    \includegraphics[width=\linewidth]{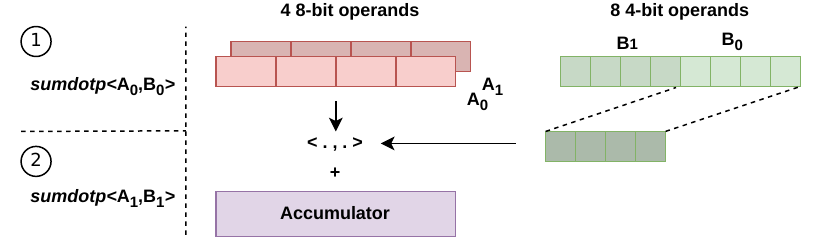}
    \caption{Execution flow of a mixed-precision \textit{sumdotp} instruction between 8-bit operand A and 4-bit operand B.}
    \label{fig:dotp-flow}
\end{figure}

To overcome this limitation and maximize the computational unit utilization, Flex-V further extends XPulpNN with mixed-precision operation support.
To efficiently enable arbitrary mixed-precision operations while avoiding the proliferation of extra instructions, Flex-V exploits the \textit{dynamic bit-scalable execution mode}: the ISA instruction only encodes the type of the operation, while the format is specified by a CSR in the core.
Figure \ref{fig:status-based} illustrates the relative decoding process: the decoder retrieves all the necessary information from the instruction and transmits it to the EX stage.
If the received op-code corresponds to a \textit{Virtual SIMD} instruction, such as a \textit{(ml)sdotp}, the decoder activates the SIMD functional unit which will execute the instruction according to CSRs values and to signals from the dedicated \textit{MAC\&Load} and \textit{Mixed-precision} controllers (MCD).
Figure \ref{fig:dotp-dp} shows the mixed-precision Dot Product (Dotp) unit.
This unit integrates a \textit{Slicer\&Router}, responsible for the extraction of the 4- and 2-bit operands from a 32-bit input word, along with the two dedicated units for the sub-byte operations.
For example, as shown in Fig.~\ref{fig:dotp-flow}, the slicer is needed in the case of a \textit{sum-of-dot-product (sdotp)} operation between an 8-bit operand A and a 4-bit operand B.
Since the single instruction can consume just four elements of the eight 4-bit words inside the register, it selects either the 16 MSBs or the 16 LSBs, according to the value of the \textit{MPC\_CNT} signal from the MCD.
Subsequently, the Router directs the desired elements to the Dotp units following the \textit{SIMD\_FMT} signal coming from the CSR, i.e. the DOTP-8b (operating on 8b inputs) for this example.

In the case of mixed-precision kernels, by re-arranging the data natively in hardware, Flex-V alleviates the substantial software overhead needed for pointer management and explicit data unpacking that would be needed otherwise.
Table \ref{tab:flex-v} shows Flex-V's performance gain over XPulpV2 and XPulpNN on dense matrix multiplication kernels with weights and activations (operands A and B in Figure \ref{fig:dotp-flow}, respectively) of different bit widths. It expresses performance in terms of MAC/cycle, isolating the inner kernel and excluding the non-idealities that arise when running complex real-world applications.
While on uniform kernels, Flex-V and XPulpNN achieve the same performance, on mixed-precision kernels, Flex-V outperforms XPulpNN by up to 6.8$\times$ and for only $5.6\%$ extra area resources.

\begin{table}[t]
    \centering
    \caption{Flex-V's performance [MAC/cycle] on MatMul kernels, against XPulpNN and XPulpV2\cite{flex_v}.}
    \begin{tabular}{c|c|c|c|c|c}
         \multicolumn{2}{c|}{\textbf{Input}}  & \textbf{XpulpV2}                                  & \textbf{XpulpNN}                                 & \textbf{Flex-V}                                    & \textbf{Flex-V}   \\ 
         \multicolumn{2}{c|}{\textbf{widths}} & \cite{kraken}                                     & \cite{gap8}                                       & \multirow{3}{*}{$\left[\frac{MAC}{cycle}\right]$} & \textbf{Speedup vs.}  \\ 
         \multicolumn{2}{c|}{[bits]}          & \multirow{2}{*}{$\left[\frac{MAC}{cycle}\right]$} & \multirow{2}{*}{$\left[\frac{MAC}{cycle}\right]$} &                                                   & \textbf{XPulpV2 / }  \\ \cline{1-2}
         \textbf{Act.}      & \textbf{Weight} &                                                   &                                                   &                                                   &  \textbf{XPulpNN}  \\ \hline
         \textbf{2}            & \textbf{2}   & -                           & 90.8                          & 91.5            &  -/$\leq1\%$ \\
         \textbf{4}            & \textbf{2}   & -                           & 7.62                          & 51.9            &  -/6.8x\\
         \textbf{4}            & \textbf{4}   & -                           & 49.5                          & 50.6            &  -/$\leq1\%$\\
         \textbf{8}            & \textbf{2}   & 4.91                        & 6.07                          & 27.8            &  5.6x/4.5x\\
         \textbf{8}            & \textbf{4}   & 6.38                        & 7.63                          & 27.6            &  4.3x/3.6x\\
         \textbf{8}            & \textbf{8}   & 16.59                       & 26.1                          & 26.9            &  1.6x/$3\%$\\
    \end{tabular}
    \label{tab:flex-v}
\end{table}

\subsubsection{IOTLB}

The cluster accesses towards the host interconnect are mediated by an IO TLB (IOTLB) unit~\cite{herov2}.
Since the Flex-V cores cannot perform virtual-to-physical address translation, the IOTLB unit is meant to ease pointer sharing with CVA6 and further prevent cluster unauthorized accesses towards the shared memory. 
The latter is a fundamental feature for critical applications: without any control, malicious or buggy applications running on the cluster could potentially cause denial-of-service to the host core or break \textit{confidentiality} (i.e., get unauthorised access to sensitive data).

The IOTLB provides 32 entries.
For each entry, CVA6 has to specify the starting and ending virtual addresses, the physical base address and the characteristics of the region: if the cluster can access it, and if it is readable or writeable.
Before offloading a task to the cluster, the host statically reserves the portions of the main memory to be shared with the cluster and then it programs the entries.
Then, once a transaction from the cluster arrives, its address is compared against the 32 virtual address ranges.
If it is within one of the available ranges and the cluster has the right permissions, the address is translated through simple subtraction of the virtual base address and the addition of the physical base address.
Otherwise, the IOTLB sends an interrupt to CVA6 to notify the cluster's attempt at accessing memory outside the expected regions.
Then, the IOTLB behaves as a simple AXI4 subordinate to not break the AXI4 protocol: for write transactions, it accepts the incoming data on the write channel, without propagating them, while for read transactions, it serves as many read beats as needed, providing an arbitrary value set at design time.
At this point, the cluster is not aware that the transaction was not allowed and continues the execution until it receives an interrupt from CVA6.
If,  in this scenario, the cluster's runtime has not been compromised by the malicious/buggy application, the cluster will gracefully interrupt its execution and resume from a known state.
If this is not the case and it is not possible to shut down the cluster, the IOTLB will anyway prevent \textit{denial-of-service} attacks and prevent unauthorised access to sensitive data.

 \section{Implementation and measurements}\label{sec:impl}
Fig.~\ref{fig:micrograph} shows the microphotograph of the Shaheen SoC, highlighting the main building blocks described in Section~\ref{sec:arch}.
The SoC is implemented in Global Foundries 22nm CMOS FD-SOI technology.
It was synthesized with Synopsys Design Compiler 2019.12, while Place \& Route was performed with Cadence Innovus 19.10.
Shaheen's 4 different clocks are generated by 4 Frequency Locked Loops (FLLs), taking a 32KHz clock in input from an off-chip ring oscillator.
The FLLs' maximum achievable output frequency at 0.8V is 600MHz.
The different peripheral PHYs (I2C, SPI, HyperBUS, ...) internally feature clock division to further scale down the input clock when needed. 

\begin{table}[t]
    \centering
    \includegraphics[width=\linewidth]{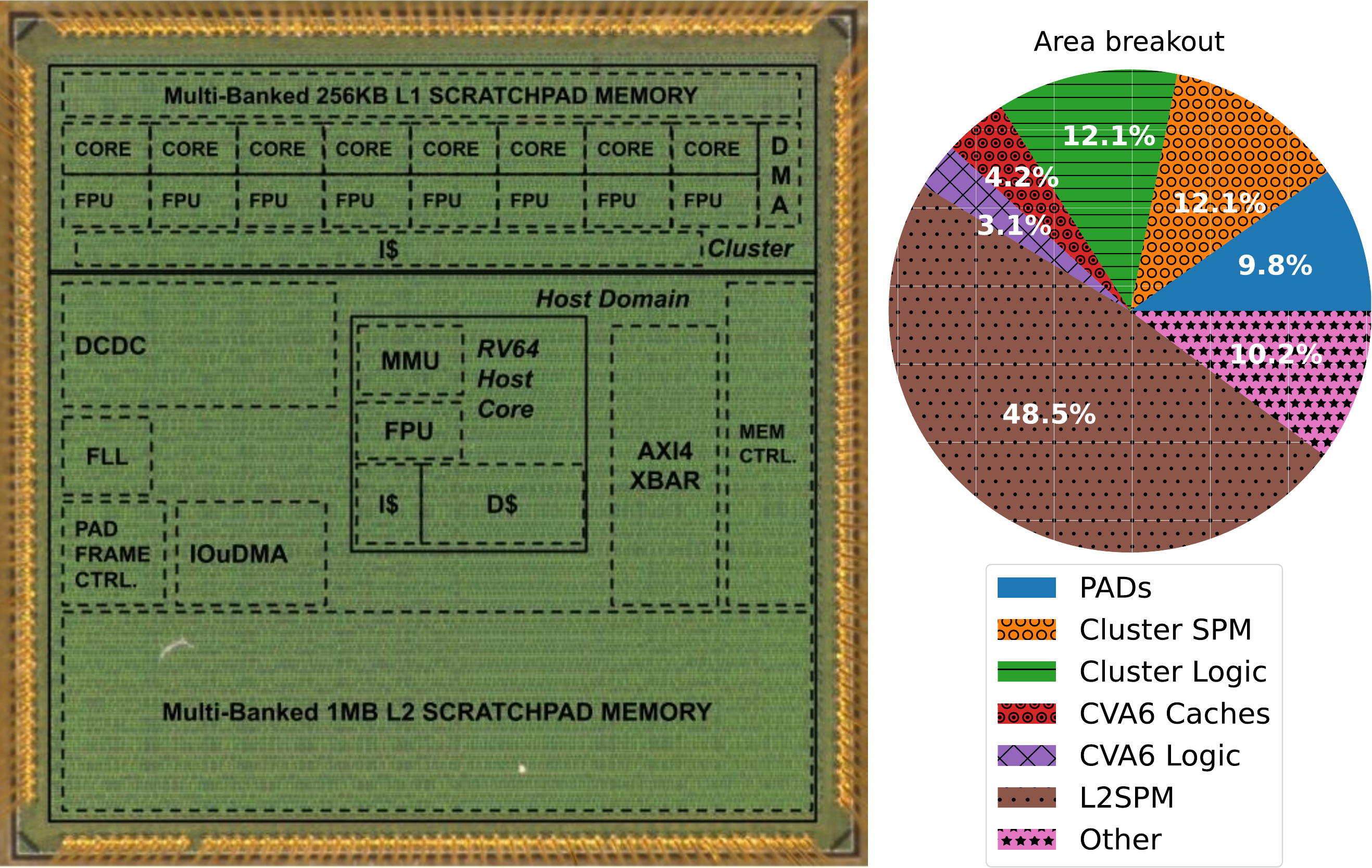}
    \captionsetup{justification=raggedright,singlelinecheck=false}
    \captionof{figure}{Die micrograph (3mm x 3mm) and area breakout.}
    \label{fig:micrograph}
    \vspace{0.4cm}
    \centering
    \begin{tabular}{cc} \hline
        Technology & CMOS22nm FD-SOI (SLVt, LVt) \\ 
        Chip Area & $9mm^2$ \\
        SRAM Memory & 1280KB \\
        VDD Range & 0.625-0.8V \\ 
        CVA6/Cluster Max Freq. & 600/500MHz \\
        Idle Power & 9-19mW (0.625-0.8V) \\ 
        Avg. Power (CVA6 Active) & 45-130mW (0.625-0.8V) \\ 
        Max. Power (Cluster Active) & 75-200mW (0.625-0.8V) \\ \hline
    \end{tabular}
    \captionsetup{justification=centering,singlelinecheck=false}
    \captionof{table}{Shaheen SoC features.}
    \label{tab:my_label}
\end{table}

\begin{figure}[t]
    \centering
    \includegraphics[width=0.45\textwidth]{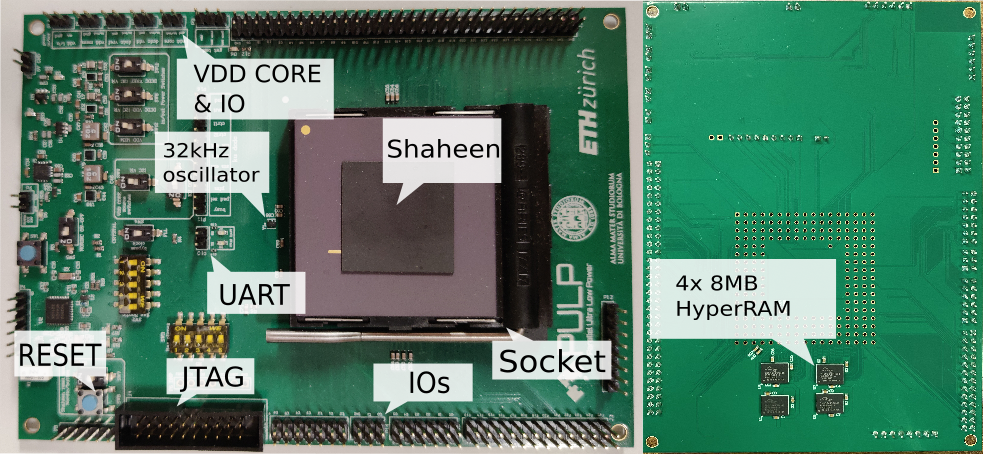}
    \caption{Shaheen test-board, top and bottom.}
    \label{fig:shaheen-test-board}
\end{figure}

Figure \ref{fig:shaheen-test-board} shows the test board developed for the bring-up and measurements. It provides four 8MB HyperRAM chips and a socket to test different chips easily. It also exposes the interfaces required to debug the chip, such as JTAG and UART, as well as pin headers connected to all the other interfaces for testing purposes. Finally, it exposes the pin headers to regulate the voltage supply of the two power domains: (i) one for the core logic and the SRAM macros, which we vary between 0.625V and 0.8V, and (ii) one for the IOs, fixed at 1.8V. While the SoC fits all the requirements for Nano-UAV navigation, the board described above has not been designed for flying, but specifically for the testing and characterization of Shaheen.

\begin{figure}[t]
    \centering
    \includegraphics[width=\linewidth]{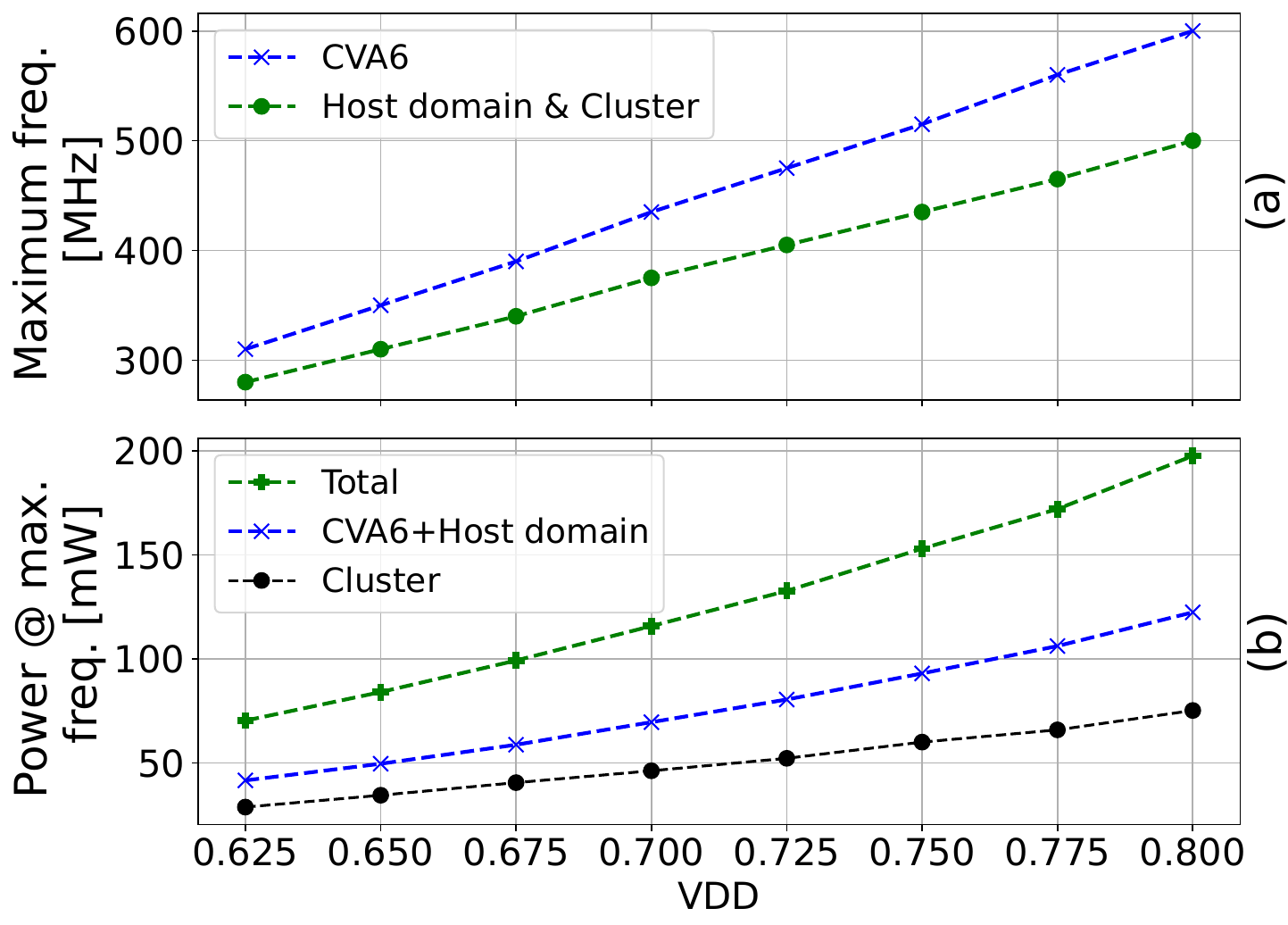}
    \caption{Maximum frequency and power envelope varying VDD.}
    \label{fig:f-sweep}
\end{figure}

First, we measure the idle power. To do so, we reduce the frequency of the SoC to 32kHz and clock gate the cluster while CVA6 is in a wait-for-interrupt state, i.e., a for loop of \textit{nop} operations. As reported in Table \ref{tab:my_label}, idle power consumption is between 9mW and 19mW, depending on the supply voltage.
Fig.~\ref{fig:f-sweep} (a) shows the measured maximum frequency varying the voltage supply of the host domain, the cluster domain, and CVA6.
The cluster and the host domain can run at up to 280MHz at 0.625V and up to 500MHz at 0.8V.
Thanks to the more aggressive pipelining, the CVA6 core can reach up to 310MHz at 0.625V and up to 600MHz at 0.8V.
Fig.~\ref{fig:f-sweep} (b) also shows the measured maximum power consumption at the highest achievable frequency for the voltage supply.
For these tests, CVA6 runs a dense FP64 matrix multiplication, and the cluster runs a dense INT32 matrix multiplication, both within an infinite loop. Then, we measure and sum the average currents consumed by the two power domains.
To get the average power consumption, we measure the power consumption of Shaheen when the cluster is clock-gated, which matches with the power consumption of CVA6, the peripheral, and the host domain together. Then, we also enable the cluster and measure the resulting total power, which coincides with the maximum power consumed by Shaheen.
Varying VDD and frequency, the power consumption of CVA6 and the host domain varies from 45mW to 130mW. On the other hand, the cluster consumes from 30mW to 70mW.

\begin{figure}[t]
    \includegraphics[width=\linewidth]{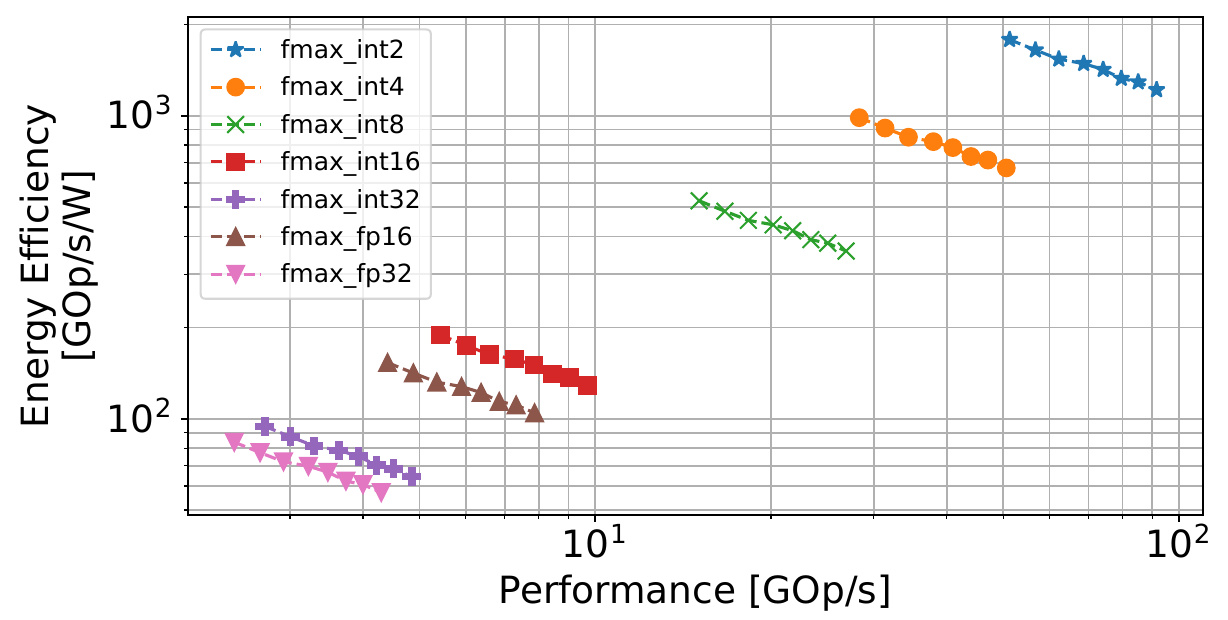}
    \caption{Pulp cluster energy efficiency on dense matrix multiplication. 1 MAC = 2 Op.}
    \label{fig:cl-en-eff}
\end{figure}

Fig.~\ref{fig:cl-en-eff} shows the cluster domain energy efficiency varying frequency, VDD and data width.
On 2-bit data, the cluster can achieve up to 90GOp/s and up to 1.8TOp/s/W.
On 8-bit data, the cluster can achieve up to 26.9GOp/s and up to 540GOp/s/W. 
All the experiments were performed running on Shaheen the various $n$-bit matrix-multiplication kernels extracted from the PULPNN library on the software-programmable cores and extracting the MAC/cycle of the inner loops, excluding the initial data arrangement overhead~\cite{flex_v}.

Lastly, we performed post-layout, parasitic annotated simulations of the HyperRAM memory controller's netlist to characterize its power consumption. At 0.8V and a speed of 1.6Gbps, it consumes 1.25mW, 70\% of which is consumed by the IOs. At 0.625V, delivering 1.1Gbps, the power consumption is 0.8mW, with 75\% of the IOs. More details about the HyperRAM controller's performance and power characterization, as well as comparisons with traditional DDR controllers, can be found in~\cite{hulk-v}.

\section{Heterogeneous Software Stack}\label{sec:swstack}

\subsection{Software stack and programming model}


Shaheen comes with a mature software stack for heterogeneous programming.
On the cluster side, we provide a lightweight bare-metal runtime that allows low programming overhead, and fast hardware functionality validation and performance profiling.
On the host side, CVA6 can either run full-ledged Buildroot-based Linux distribution (v5.16.9) on top of the Bao Hypervisor \cite{bruno_h} or a bare-metal runtime, and both are equipped with a dedicated driver for the cluster management.
The APIs provided by the cluster runtime and the CVA6's driver are already sufficient to run heterogeneous code on the platform.
However, one must write two different codes for the host and cluster.
To avoid this, Shaheen adapts the OpenMP 5 framework from HERO~\cite{herov2}, allowing users to use a high-level, directive-based, intuitive programming interface to efficiently offload the computationally intensive part of a program to the cluster within one single heterogeneous source code. 
Also, to map the execution of QNNs on the cluster, we adopt the data and execution flow presented in Dory\cite{dory}. 
Dory is a tool that given the description of a QNN in input generates the corresponding C code to be executed on parallel programmable clusters.
Dory calculates data tiling solutions fitting the available L1SPM (where it puts the data to be processed by the cluster) and it schedules the DMA data transfers from the main memory to the L1SPM and vice-versa.
Thanks to the efficiency of tiling and double-buffering, when the execution is not memory bound, data movements overlap with computation for more than $95\%$ of the execution time\cite{dory}.

\subsection{Offload mechanism \& Performance}

To perform the offload, CVA6 \textit{lazily} (at first occurrence) loads the cluster code into the L2SPM to then communicate to the cluster where is the code to execute.
Such a mechanism requires a few thousand clock cycles, depending on the length of the code.
Hence, when the cluster execution time is very short ($<$100k cycles), the cluster's offload overhead (i.e., loading the code) dominates the total execution time and reduces the speedup. Based on our (empirical) experience, this is a very uncommon case.

\begin{figure}[t]
    \centering
    \includegraphics[width=\linewidth]{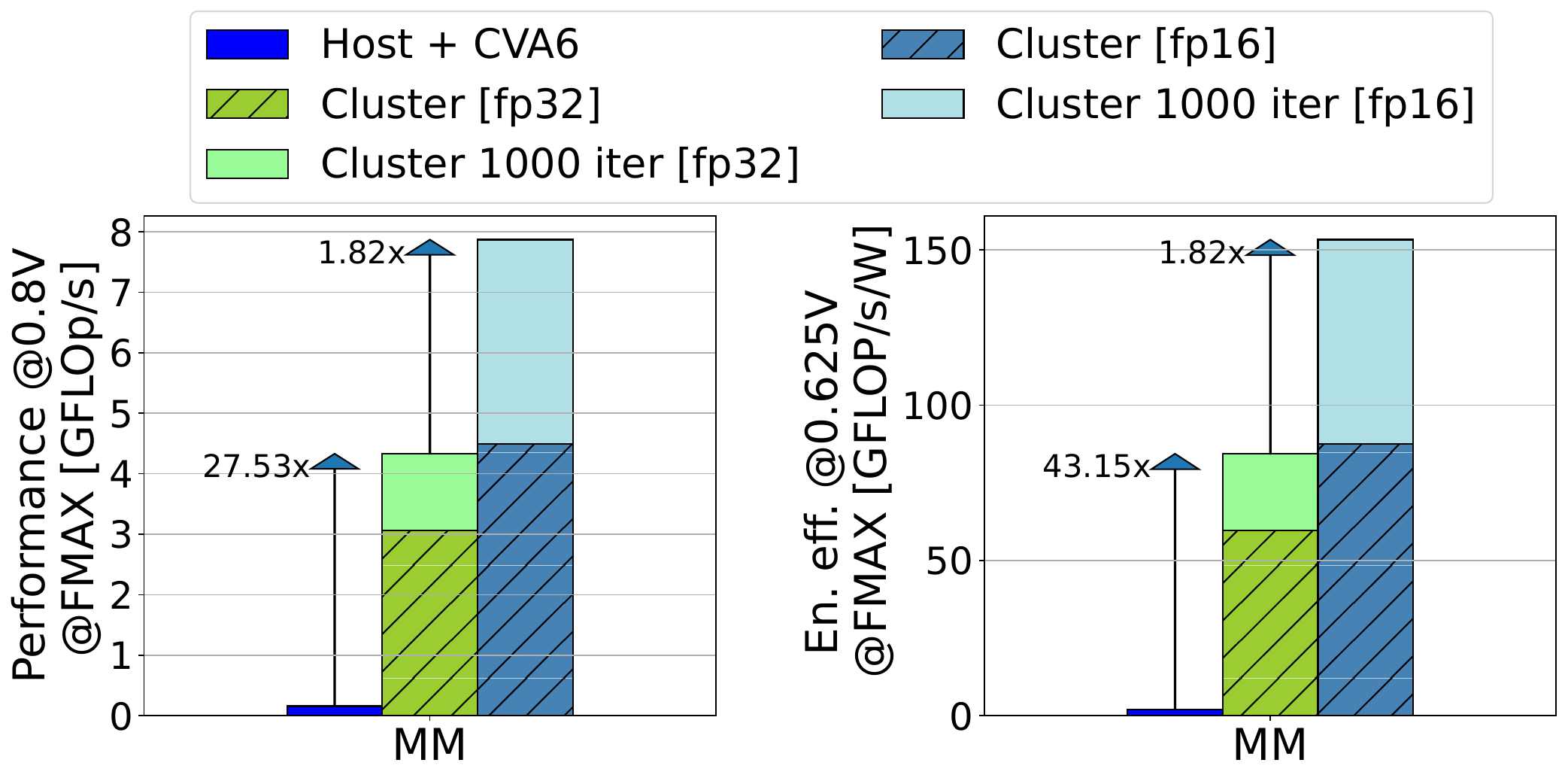}
    \caption{Offload performance breakout on an FP MM.}
    \label{fig:oflload_overhead}
\end{figure}

Figure \ref{fig:oflload_overhead} shows the offload speedup and overhead over an FP matrix multiplication.
It is important to notice how the cluster can run such a benchmark with reduced precision (down to FP16), exploiting the SIMD extensions otherwise unavailable on the CVA6 core.

The plot on the left in Fig.~\ref{fig:oflload_overhead} shows CVA6 and cluster performance at the maximum frequency at 0.8V (600MHz for CVA6, 500MHz for the cluster), and it also highlights the speed up.
The figure shows the acceleration when executing the accelerated kernel once or 1000 times on the cluster; the first case represents low code utilization, while the second represents high code utilization.
In each execution, the cluster performs the computation on a different couple of input matrices. At the same time, it fetches the input matrices for the next execution and writes back the result of the previous computation. Data movement is performed through the DMA and overlaps computation.
On a simple FP32 matrix multiplication, the cluster can deliver up to 4.3 GFLOp/s, which is roughly 27 times more than CVA6.
Furthermore, the plot shows once again the benefit of scaling down the number precision, which is not a possibility on CVA6. 

The plot on the right in Figure~\ref{fig:oflload_overhead} compares the energy efficiency achieved by CVA6 and the host domain against the cluster on the same benchmark, with the IPs working at the maximum frequency at 0.65V (280MHz for the cluster and 310MHz for CVA6). 
On the reduced-precision matrix multiplication, the cluster can reach up to 157 GOp/s/W, while CVA6 can only provide 2 GOp/s/W, $\approx 80 \times$ less. 

\section{Benchmarking}\label{sec:bench}

\begin{figure}[t]
    \centering
    \includegraphics[width=\linewidth]{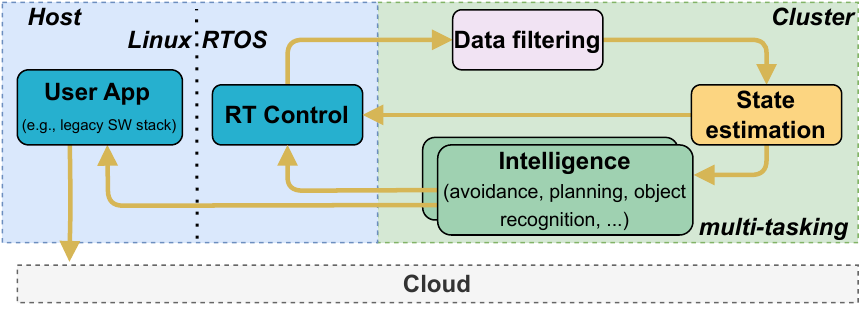}
    \caption{Shaheen's autonomous mission \& flight functional blocks.}
    \label{fig:control-loop}
\end{figure}

Figure \ref{fig:control-loop} presents the loop that Shaheen executes to achieve autonomous flight while executing other auxiliary tasks, and how it maps on Shaheen's hardware.
In the first instance, it collects and filters the sensors' input data, which are subsequently used to estimate the current state.
Then, in the ``intelligence" block it has to independently determine the next state (i.e., what to do next) and carry out the target auxiliary tasks such as object detection, recognition or monitoring \cite{cereda2023deep,fft_uav}.
Once the next state is determined, the control part actuates the change.
In Shaheen, the first three phases (filtering, state estimation, and intelligence) are mapped on the cluster hardware, while the real-time control is left to the host core, also executing the general-purpose OS, to leverage its legacy software stack for non-real-time tasks (e.g. transmitting the classification of a detected object to the cloud through the legacy network stack).
In this section, we focus on benchmarking the cluster on its target tasks as they are the most computing-intensive phases and potential bottlenecks of the autonomous flight and mission loop.

Filtering of the input data and state estimation as well as intelligence tasks like path planning or structural health monitoring usually rely on general-purpose DSP primitives\cite{wavelet_drones,fft_uav}.
At the same time, QNN inference is widely adopted for tasks like classification or recognition for obstacle avoidance or object recognition and localization\cite{cereda2023deep,dronet,agilicious,howard2017mobilenets,hawq}. 
However, one limitation of the QNN inference at the edge is the mainstream adoption of the \textit{train-once-deploy-everywhere} approach, which trains the networks offline and then deploys them later on the embedded devices, where no further modifications to the weights happen.
This approach prevents the models to adapt in the deployment environment and possibly leads to accuracy degradation and unreliability \cite{amodei2016concrete}.
On-device learning potentially overcomes this limitation by enabling small portions of the training to happen on the field, directly on the MCU~\cite{trainlib_git}.
Thus, we benchmark the proposed SoC on three sets of kernels representative of the different tasks described above, i.e., (i) general-purpose DSP, (ii) DNN, and (iii) online learning benchmarks.

\subsection{General Purpose DSP}

\begin{figure}[t]
    \centering
    \includegraphics[width=\linewidth]{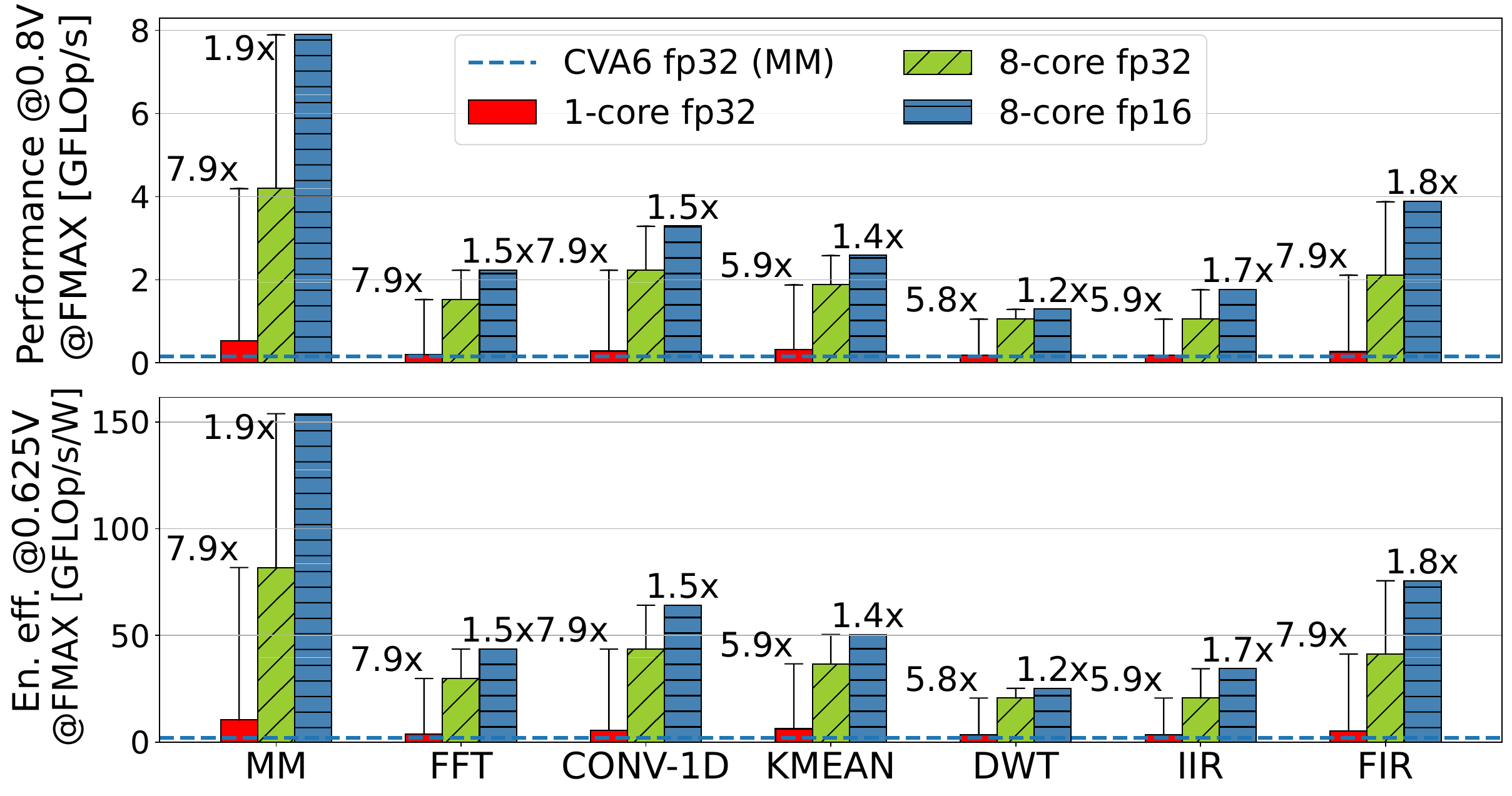}
    \caption{Performance and en. eff. delivered by the cluster on general-purpose DSP benchmarks.}
    \label{fig:cl_dsp}
\end{figure}

Figure \ref{fig:cl_dsp} shows the cluster performance and energy efficiency over seven open-source FP benchmarks\cite{dsp_git} representative of DSP applications for filtering, feature extraction classification, and basic linear algebra functions, relevant both for input data filtering, state estimation but also intelligence tasks such as path planning or structural health monitoring\cite{wavelet_drones,fft_uav}.
To show the advantage given by parallelism and reduced-precision computation, such benchmarks are executed both at full precision (FP32) on a single Flex-V core, and then on 8 cores at full and lower precision (FP16 \& bfloat16) to exploit the available packed-SIMD support.

Some of the benchmark kernels are representative of digital data acquisition and analysis, such as the Finite Impulse Response (FIR) and Infinite Impulse Response (IIR) filters. 
To characterize the cluster on frequency-domain applications, we run a decimation-in-frequency radix-2 variant of the Fast Fourier Transform (FFT) and a Discrete Wavelet Transform (DWT), a standard kernel used for feature extraction.
As for more state-estimation-oriented kernels, we provide the performance results when executing a K-Means classifier kernel.
Lastly, we also benchmark two classical basic linear algebra kernels such as a Matrix Multiplication and a 1D Convolution.

As the plots in Fig.~\ref{fig:cl_dsp} show, on all these benchmarks, thanks to the ISA extension not available on the host core, a single Flex-V running at 500MHz provides from 1$\times$ to 3$\times$ the performance delivered by CVA6 on a dense and regular matrix multiplication at 600MHz.
Furthermore, the parallel execution of the benchmarks on 8 Flex-V cores can give an additional speed-up between 5.9 and 7.9 times when compared to single-core execution.
Leveraging the reduced precision arithmetic can further provide almost a 2x speed-up, allowing the core to reach up to 7.9GFLOp/s and up to 157 GFLOp/s/W.

\subsection{QNN inference}

In this subsection, we focus on two real-world 8-bit QNNs fine-tuned for Nano-UAVs application scenarios, namely Tiny-PULP-Dronet~\cite{dronet} and FrontNet~\cite{cereda2023deep}, as well as two aggressively quantized mixed-precision QNNs for object detection and classification.
The Tiny-PULP-Dronet is a lightweigth QNN based on the ResNet architecture \cite{hawq} and it enables autonomous navigation within tight spaces avoiding obstacle collision.
FrontNet on the other hand is based on the MobileNet\cite{howard2017mobilenets} architecture and it is used for Human-Robot Interaction (HRI): it allows the nano-drone to recognize a face and follow it.
The cluster is able to achieve 320FPS on a Tiny-PULP-Dronet and 260FPS on an optimized 6.7MMAC FrontNet, which is well above the 20FPS needed to achieve autonomous flight \cite{dronet,cereda2023deep}.
This means that more than $90\%$ of the cluster's computational capabilities are actually available to carry out other activities.

\begin{table}[t]
    \centering
    \caption{Accuracy, memory footprint, perf. \& en. of end-to-end networks.}
    \label{tab:benchmark_nets}
    \begin{tabular}{c|c|c|c}
        \hline
        \textbf{\multirow{2}{*}{Network}}    & \textbf{MobilNetV1} & \textbf{MobileNetV1}      & \textbf{ResNet20} \\
                                             & \textbf{(8b)}       & \textbf{(8b4b)}            & \textbf{(4b2b)} \\
        \hline
        \textbf{Top-1 Acc.}       & 69.3\%             & 66.0\% & 90.2\%\\
        \hline
        \textbf{Deg. w.r.t. 8b}   & -                  & 3.3\%    & 0.15\%\\
        \hline
        \textbf{Model size}       & 1.9 MB             & 997 kB            & 142 kB \\
        \hline
        \textbf{MACs}             & 325M                & 328M               & 10M \\
        \hline
        \textbf{Mem. saved}       &                    & 47\%              & 63\%\\
        \hline
        \multicolumn{4}{c}{\textbf{Avg. Perf. : Latency : Energy [MAC/cycle : ms : mJ]}} \\
        \multicolumn{4}{c}{\textbf{@0.8V -run to sleep-}} \\
        \hline
        \textbf{XpulpV2}         & 5.6 :      & 3.2 :       & 4.8 :      \\
        \textbf{GAP9, 450MHz}   & 126 : 5.87 & 227 : 10.23 & 4.9 : 0.22 \\
        \hline
        \textbf{XpulpNN}         & 6.0 :     & 2.7 :       & 4.4 : \\
        \textbf{Kraken, 380MHz}  & 141 : 12.7 & 319 : 28.76 & 6.4 : 0.58 \\
        \hline
        \textbf{Flex-V}          & 6.0 :      & 5.8 :       & 11.2 : \\
        \textbf{Shaheen, 500MHz} & 108 : 8.55 & 119 : 8.83  & 1.9 : 0.15 \\
        \hline
        \end{tabular}
\end{table}

Stemming the analysis from the QNNs mentioned below, we first benchmark the cluster on a relatively big (325MMAC) 8-bit MobileNetV1\cite{howard2017mobilenets} for object classification.
Then, we extend the analysis to a mixed-precision MobileNetV1 with 8-bit activations and 4-bit (8b4b) weights and an aggressively quantized 4b2b ResNet-20\cite{he2015deep} for object detection.
The two MobileNetV1 networks have been trained on ImageNet while the 4b2b ResNet-20 targets CIFAR10.
As table \ref{tab:benchmark_nets} shows, reducing the operands' precision does not automatically jeopardize the accuracy: in the case of the MobileNetV1, there is a 47\% memory footprint reduction for a negligible 3\% accuracy loss, from 69\% to 66\%~\cite{flex_v}, while the ResNet-20 achieves $90.2\%$ accuracy \cite{hawq}.
As shown in table \ref{tab:benchmark_nets}, Flex-V is the only version of Ri5cy able to efficiently deal with mixed precision networks: in terms of MAC/cycles, on the 4b-2b mixed-precision ResNet-20, it achieves 2.3$\times$ and 2.5$\times$ of speedup with respect to XPulpNN and XPulpV2.
Table \ref{tab:benchmark_nets} also compares the latency and energy consumed by Shaheen over the three networks when running at the maximum frequency at 0.8V, compared to two other 8-core clusters respectively implementing the baseline XPulpV2 instructions or the XPulpNN, namely GAP9\cite{gap8} and Kraken\cite{kraken}.
On the uniform-precision MobileNetV1, thanks to the higher frequency and optimized ISA, Shaheen's cluster provides the smallest latency, but not the lowest energy, due to higher power consumption (70mW) when compared to GAP9 (50mW), being the latter tuned for energy-efficient operation.
As soon as the mixed-precision extensions can be exploited, Shaheen's cluster emerges both as the fastest and most energy-efficient one.

\subsection{Online training}

\begin{figure}[t]
    \centering
    \includegraphics[width=\linewidth]{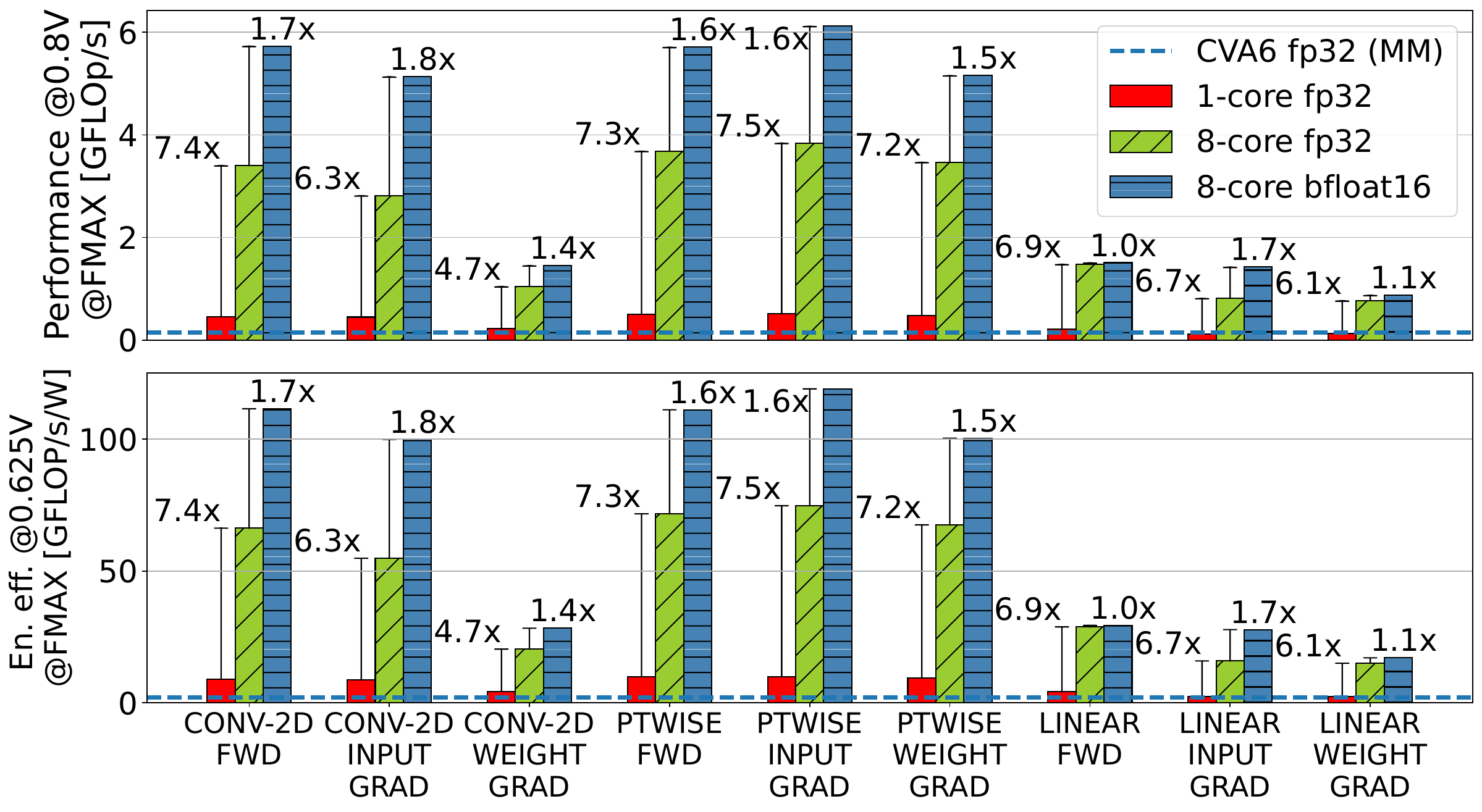}
    \caption{Performance and En. Eff. delivered by the cluster on FP training benchmarks.}
    \label{fig:cl_fp}
\end{figure}

\begin{table*}[t]
\centering
\caption{Comparison with SoA SoCs.}
\label{fig:comp_soa}
\begin{tabular}{|c|c|c|c|c|c|c|c|}
\hline
\multirow{2}{*}{} &
  \multirow{2}{*}{\textit{\begin{tabular}[c]{@{}c@{}}Neo - Cheshire\\ \cite{cheshire}\end{tabular}}} &
  \multirow{2}{*}{\textit{\begin{tabular}[c]{@{}c@{}}Jia et al.\\ \cite{esp_tapeout}\end{tabular}}} &
  \multirow{2}{*}{\textit{\begin{tabular}[c]{@{}c@{}}STM32-H7\\ \cite{stm32h7}\end{tabular}}} &
  \multirow{2}{*}{\textit{\begin{tabular}[c]{@{}c@{}}Ju et al.\\ \cite{Ju}\end{tabular}}} &
  \multirow{2}{*}{\textit{\begin{tabular}[c]{@{}c@{}}GAP9\\ \cite{gap8}\end{tabular}}} &
  \multirow{2}{*}{\textit{\begin{tabular}[c]{@{}c@{}}Kraken \\ \cite{kraken}\end{tabular}}} &
  \multirow{2}{*}{\textit{Shaheen (ours)}} \\
                    &            &            &            &            &          &            &            \\ \hline
\textit{Target}     & Prototype  & Prototype  & Product    & Prototype  & Product  & Prototype  & Prototype  \\ \hline
\textit{Technology} & 65nm       & 12nm       & 40nm       & 65nm       & 22nm     & 22nm FDSOI & 22nm FDSOI \\ \hline
\textit{Die Size}   & $6,4mm^2$  & $21,6mm^2$ & -          & $4,47mm^2$ & $12mm^2$ & $9mm^2$    & $9mm^2$    \\ \hline
\multirow{3}{*}{\textit{\begin{tabular}[c]{@{}c@{}}CPU / \\ ISA\end{tabular}}} &
  \multirow{3}{*}{\begin{tabular}[c]{@{}c@{}}CVA6/ \\ RV64GC\end{tabular}} &
  \multirow{3}{*}{\begin{tabular}[c]{@{}c@{}}4xCVA6 / \\ RV64GC\end{tabular}} &
  \multirow{3}{*}{\begin{tabular}[c]{@{}c@{}}Cortex-M7 + \\ CortexM4 / \\ ARM32\end{tabular}} &
  \multirow{3}{*}{\begin{tabular}[c]{@{}c@{}}10xRISC-V / \\ RV32\end{tabular}} &
  \multirow{3}{*}{\begin{tabular}[c]{@{}c@{}}10x RI5CY / \\ RV32\end{tabular}} &
  \multirow{3}{*}{\begin{tabular}[c]{@{}c@{}}9x RI5CY-XNN / \\ RV32\end{tabular}} &
  \multirow{3}{*}{\begin{tabular}[c]{@{}c@{}}CVA6 + \\ 8x FLEX-V / \\ RV64GC + RV32\end{tabular}} \\
            &        &            &            &            &          &            &            \\
            &        &            &            &            &          &            &            \\ \hline
\multirow{1}{*}{\textit{Power Env.}} &
  \multirow{1}{*}{300mW} &
  \multirow{1}{*}{1.83W} &
  \multirow{1}{*}{-} &
  \multirow{1}{*}{116mW-589mW} &
  \multirow{1}{*}{50mW} &
  \multirow{1}{*}{300mW} &
  \multirow{1}{*}{195mW} \\ \hline
\textit{Max. Freq.}   & 300MHz & 1,5GHz     & 480-240MHz & 400MHz     & 450MHz   & 330-390MHz & 500-600MHz \\ \hline
\multirow{4}{*}{\textit{\begin{tabular}[c]{@{}c@{}}Memory \\ exposed to \\ the CPU\end{tabular}}} &
  \multirow{4}{*}{1GB RPC} &
  \multirow{4}{*}{2GB SerDes} &
  \multirow{4}{*}{\begin{tabular}[c]{@{}c@{}}512kB On-chip \\ SRAM\end{tabular}} &
  \multirow{4}{*}{\begin{tabular}[c]{@{}c@{}}150kB On-chip \\ SRAM\end{tabular}} &
  \multirow{4}{*}{\begin{tabular}[c]{@{}c@{}}1.5MB On-chip \\ SRAM + \\ 8-64MB \\ HyperBUS (XIP)\end{tabular}} &
  \multirow{4}{*}{\begin{tabular}[c]{@{}c@{}}1.5MB On-chip \\ SRAM\end{tabular}} &
  \multirow{4}{*}{\begin{tabular}[c]{@{}c@{}}1MB On-chip \\ SRAM + \\ 32-512MB \\ HyperBUS\end{tabular}} \\
                   & &            &            &            &          &            &            \\
                   & &            &            &            &          &            &            \\
                   & &            &            &            &          &            &            \\ \hline
\multirow{2}{*}{\textit{\begin{tabular}[c]{@{}c@{}}Supported\\ OS\end{tabular}}} &
  \multirow{2}{*}{\begin{tabular}[c]{@{}c@{}}Linux / \\ RTOS\end{tabular}} &
  \multirow{2}{*}{\begin{tabular}[c]{@{}c@{}}Linux / \\ RTOS\end{tabular}} &
  \multirow{2}{*}{RTOS} &
  \multirow{2}{*}{RTOS} &
  \multirow{2}{*}{RTOS} &
  \multirow{2}{*}{RTOS} &
  \multirow{2}{*}{\begin{tabular}[c]{@{}c@{}}Linux + \\ RTOS \end{tabular}} \\
                &    &            &            &            &          &            &            \\ \hline
\multirow{2}{*}{\textit{\begin{tabular}[c]{@{}c@{}}Security \\ Features\end{tabular}}} &
  \multirow{2}{*}{-} &
  \multirow{2}{*}{-} &
  \multirow{2}{*}{\begin{tabular}[c]{@{}c@{}}Crypto/hash \\ processor\end{tabular}} &
  \multirow{2}{*}{-} &
  \multirow{2}{*}{\begin{tabular}[c]{@{}c@{}}AES128/256 acc., \\ PUF\end{tabular}} &
  \multirow{2}{*}{-} &
  \multirow{2}{*}{\begin{tabular}[c]{@{}c@{}}fence.t, IOTLB, \\ PMP, Hypervisor\end{tabular}} \\
              &      &            &            &            &          &            &            \\ \hline
\multirow{3}{*}{\textit{\begin{tabular}[c]{@{}c@{}}SW INT/FP \\ arithmetic \\ support\end{tabular}}} &
  \multirow{3}{*}{SP/DP-FP} &
  \multirow{3}{*}{SP/DP-FP} &
  \multirow{3}{*}{SP-FP} &
  \multirow{3}{*}{-} &
  \multirow{3}{*}{\begin{tabular}[c]{@{}c@{}}bfloat16, fp16/32, \\ int32/16/8\end{tabular}} &
  \multirow{3}{*}{\begin{tabular}[c]{@{}c@{}}bfloat16, fp16/32, \\ int32/16/8/4/2\end{tabular}} &
  \multirow{3}{*}{\begin{tabular}[c]{@{}c@{}}bfloat16, fp16/32, \\ int32/16,\\ int-mixed8/4/2\end{tabular}} \\
                &    &            &            &            &          &            &            \\
                &    &            &            &            &          &            &            \\ \hline
\multirow{3}{*}{\textit{\begin{tabular}[c]{@{}c@{}}Peak SW INT \\ Performance\end{tabular}}} &
  \multirow{3}{*}{\begin{tabular}[c]{@{}c@{}}$0.5$GOp/s \\ (32b)\end{tabular}} &
  \multirow{3}{*}{\begin{tabular}[c]{@{}c@{}}$1.5$GOp/s \\ (32b)\end{tabular}} &
  \multirow{3}{*}{\begin{tabular}[c]{@{}c@{}}390MOp/s\\ (8b)\end{tabular}} &
  \multirow{3}{*}{\begin{tabular}[c]{@{}c@{}}16GOp/s\\ (8b)\end{tabular}} &
  \multirow{3}{*}{\begin{tabular}[c]{@{}c@{}}15.6GOp/s\\ (8b)\end{tabular}} &
  \multirow{3}{*}{\begin{tabular}[c]{@{}c@{}}22GOp/s(8b-8b) \\ 45GOp/s(4b-4b) \\ 85GOp/s(2b-2b)\end{tabular}} &
  \multirow{3}{*}{\begin{tabular}[c]{@{}c@{}}26GOp/s(8b-8/4/2b) \\ 50GOp/s(4b-4/2b) \\ 90 GOp/s(2b-2b)\end{tabular}} \\
                &    &            &            &            &          &            &            \\
                &    &            &            &            &          &            &            \\ \hline
\multirow{2}{*}{\textit{\begin{tabular}[c]{@{}c@{}}Peak SW FP \\ Performance\end{tabular}}} &
  \multirow{2}{*}{\begin{tabular}[c]{@{}c@{}}$0.5$GFLOp/s \\ (32b)\end{tabular}} &
  \multirow{2}{*}{\begin{tabular}[c]{@{}c@{}}$1.5$GFLOp/s \\ (32b)\end{tabular}} &
  \multirow{2}{*}{\begin{tabular}[c]{@{}c@{}}240MFLOp/s \\ (32b)\end{tabular}} &
  \multirow{2}{*}{-} &
  \multirow{2}{*}{\begin{tabular}[c]{@{}c@{}}3.3GFLOp/s\\ (32b)\end{tabular}} &
  \multirow{2}{*}{\begin{tabular}[c]{@{}c@{}}3.12GFLOp/s\\ (32b)\end{tabular}} &
  \multirow{2}{*}{\begin{tabular}[c]{@{}c@{}}4.0GFLOp/s(32b) \\ 7.9GFLOp/s(16b)\end{tabular}} \\
            &        &            &            &            &          &            &            \\ \hline
\end{tabular}
\end{table*}

In this subsection, we benchmark the cluster against a set of open-source kernels to enable online learning on MCU controllers~\cite{trainlib_git}.
In particular, we benchmark three very popular layers such as 2D Convolution, Pointwise and Fully-Connected which are the three building blocks of Convolutional NNs (CNNs), used to find patterns in images.
Convolutional and pointwise layers are the core building blocks of CNN, where most of the computation happens, and are used to perform feature extraction.
The Fully connected layer connects the information extracted from the previous steps (i.e., Convolution layer and Pooling layers) to the output layer and eventually classifies the input into the desired label.
For each layer, we consider the three phases of training: (i) the forward pass, to compute the output result and hence the loss, (ii) the backward computation of the gradients with respect to the activations, and (iii) the backward computation of the gradients with respect to the weights.
The kernels we leverage map each of these computation phases directly to one matrix multiplication containing all the matrix multiplications needed to obtain the output\cite{trainlib_git}. Depending on the matrices' shapes, the amount of parallelizable work changes, and hence the performance\cite{Amdahl}.
Figure \ref{fig:cl_fp} shows performance and energy efficiency over such benchmarks.
As for the DSP benchmarks, the parallelization provides a significant speed-up for most of them.
Except for the weight gradient computation on the convolution kernel, which achieves a 4.7$\times$ speedup, the parallelization provides between 6.1x and 7.5x faster execution.
At the same time, leveraging the bfloat16 format (providing a wide dynamic range explicitly thought for ML training) and the dedicated SIMD extensions provides up to 1.8x more performance. 
Overall, the cluster is able to achieve up 6.2 GFLOp/s and 120 GFLOp/s/W on this class of benchmarks.

\section{Comparison with State-of-the-Art}\label{sec:comp_silicon}

Table \ref{fig:comp_soa} shows Shaheen against 6 SoCs for UAVs, both from industry and academia.
To have a thorough comparison, we extend it also with SoC not explicitly optimized for UAVs but with similar general-purpose software performance and functionalities that could fit the purpose, namely Cheshire \cite{cheshire}, the work from Jia et.al. \cite{esp_tapeout}, the STM32-H7\cite{stm32h7} and the work by Ju et.al. \cite{Ju}.
From an architectural viewpoint, Ju et. al. \cite{Ju} consists of a homogeneous systolic array of RV32 cores, while Jia et al. \cite{esp_tapeout} instantiates a cluster of four RV64 cores along with a set of hardwired ASIC accelerators.
More advanced nano-UAV SoCs, such as GAP9\cite{gap8} and Kraken\cite{kraken}, incorporate an RV32 CPU that can offload compute-intensive tasks to a parallel cluster of cores with the same ISA.

In this context, Shaheen is the first silicon demonstrator of a heterogeneous RV64/RV32 architecture.
When offloading compute-intensive tasks to the fully-programmable parallel cluster of Flex-V cores, performance can be improved by up to 2 orders of magnitude achieving state-of-the-art performance with up to 90 GOp/s on heavily quantized integer tasks and up to 7.9 GFLOp/s/W on 16-bit floating point tasks.
Shaheen stands out as the only nano-UAV SoC that provides Linux, hypervisor, and security capabilities to the host enabling the secure co-existence of user applications running on full-fledged OSes and control tasks running on real-time OSes while providing up to 512MB of low-cost and low-power off-chip memory within the power envelope of 200 mW.

%
%
%

\section{Conclusion}\label{sec:concl}

We presented Shaheen: a heterogeneous and flexible SoC implemented in \SI{22}{\nano\meter} FDX technology.
Shaheen features a Linux-capable RV64 core, compliant with the v1.0 ratified Hypervisor extension. 
To the best of our knowledge, it is the first silicon implementation fully compliant with the ratified RISC-V ISA Hypervisor extension. 
It features support for timing channel protection to isolate concurrent execution of multiple software stacks (trusted and untrusted), preventing security threats and ensuring multi-domain operations. 
It provides up to 512MB of main off-chip HyperRAM memory, large enough to host general-purpose OSs as well as RTOSs.
Also, it is the first silicon implementation of a heterogeneous MCU coupling an RV64 host together with a multi-core RV32 cluster, achieving up to 90GOp/s and up to 1.8TOp/s/W on 2-bit integer kernel and up to 26.9GOp/s and up to 540GOp/s/W on 8-bit integer kernels. 

After this thorough evaluation, we envision the miniaturization of the testing PCB (see Fig. \ref{fig:shaheen-test-board}) and development of \textit{ad-hoc} control software, tightly coupled with the physical characteristics of the board, to achieve real-world nano-UAV flight, exploiting Shaheen's secure and scalable architecture with host/cluster decoupling and advanced virtualization.
Overall, Shaheen is the first prototype SoC providing support for general-purpose OSs within a $200mW$ power envelope while offering state-of-the-art performance over a wide spectrum of applications, thanks to the programmable multi-core cluster.
All the IPs integrated within Shaheen are released as open source\footnote{\url{https://github.com/pulp-platform}} under a liberal license to foster future research in the area of AI-IoT computing devices.

\bibliographystyle{IEEEtran}
\bibliography{IEEEabrv,main}

\end{document}